\documentclass[12pt]{article}
\usepackage{graphicx,amsmath,amssymb,cite}
\usepackage{epsfig,epsf}
\usepackage{graphicx}
\usepackage{epstopdf}
\newcommand{\dif}{\mathrm{d}}

\begin{document}
%\maketitle

\begin{center}
{\Large{\bf The $J/\psi$ Way to Nuclear Structure }} \\
[0.5cm]

{\large A. Caldwell~$^1$, and H. Kowalski~$^2$ } \\ [0.3cm]

{\it $^1$ Max Planck Institute for Physics, M\"unchen}
\\[0.1cm]

{\it $^2$ Deutsches Elektronen-Synchrotron DESY, D-22607 Hamburg, Germany}\\[0.1cm]

\end{center}

\begin{center}
{\bf Abstract}
\end{center}
 
We propose to investigate the properties of nuclear matter by measuring the elastic scattering of $J/\psi$ on nuclei with high precision. The $J/\psi$ mesons are produced from the photons emitted in high energy electron-proton or electron-nucleus scattering in the low-x region. The measurement could be performed at the future ENC, EIC or LHeC  facilities.    
\vspace*{3 cm}

\section{Introduction}
Nuclei have  fascinating properties which are very interesting for basic and applied science. A better understanding of nuclei could lead to efficient energy production and other industrial uses.  The basic models of the nuclei - the liquid drop and the shell model - give a simple but refined understanding of many nuclear properties. They are built on the fact that protons and neutron in nuclei behave like an incompressible fluid and are arranged in  highly ordered quantum states, or shells. A familiar but very puzzling property of  nuclear matter is the growth of its volume proportionally to the number of nucleons. Although often explained by the Fermi nature of the nucleons it is mysterious when viewed from the particle physics perspective because nucleons consist of quarks and gluons which are point-like and nearly massless\footnote{The size of the hadrons made of quarks and gluons is also not understood.}. We note that the sizes of atoms  are nearly independent of the charge number Z.      

One of the main reasons why, after almost 80 years of investigations of nuclear structure, some of the basic properties are so poorly understood is a lack of proper tools to view  inside nuclei.
In the past, the most direct information about nuclear structure was obtained by scattering electron beams off nuclei. The electron is a very good probe because it penetrates into the nuclear interior without being absorbed. Unfortunately, electrons can only see the electric charge distributions. They  are not sensitive to the distribution of strong fields which keep nuclei together.  Another important source of information comes from the scattering of low energy protons on nuclei. The protons sense the full matter distribution but they are themselves complicated objects whose interactions with matter are not very  well known. 
%The strong interactions are only understood at distances which are much smaller than the proton size, where  perturbative QCD is valid.  
Our understanding of strong interactions is currently only robust at distances much smaller than the proton size, where perturbative QCD is applicable. Although Lattice Gauge 
Theory has firmly established that QCD is the correct theory of  strong interactions at large distances,  its applications  to  hadronic interactions are only now being developed~\cite{Ishii:2006ec,Beane:2008ia,Wilczek}.
%For recent promising work in this direction, we refer the reader to Refs.." 

In recent years HERA has shown that, in deep inelastic electron-proton scattering at small $x$,  up to 20\% of events are of diffractive origin\footnote{Here diffraction means that the proton stays intact.}.  This is 
comparable to the rate of the elastic reactions in hadron-hadron scattering at high energy. 
In hadronic reactions the rate of elastic and inelastic events 
is connected  by the optical theorem. Before HERA, the optical theorem was not expected to play any role in  electron-proton scattering because the elastic $ep$ component is miniscule. The abundance of the diffractive (i.e. quasi elastic) processes seen at HERA indicates therefore that some intermediate states have to be formed to which the optical theorem can be applied.  
The theoretical framework which describes naturally the formation of such  intermediate states  is the dipole picture.  Dipoles are small quark-antiquark pairs which interact by color exchange.
In contrast to hadron scattering, where the projectiles are complicated, the color dipoles  are simple objects. Their creation, interaction  and annihilation  has a simple  description in  perturbative QCD combined with QED. 
%In this picture the optical theorem allows to relate the absorptive 
%and elastic scattering processes. 
A multitude of HERA measurements  such as inclusive cross-sections,  charm production, diffraction,  jet production, vector-meson production and DVCS cross sections are
accurately explained within the dipole picture\cite{
Nemchik1996,Gotsman1995,Dosch1996,GBW,BGK,MSM,CS,Forshaw2003,
Frankfurt2005,KLMV} .

A particular final state into which a dipole transforms acts as an analyzer. For example, jets select very small dipoles, $J/\psi$ production selects charm anti-charm  dipoles, $\rho$ mesons  $u\bar u$ or $d\bar d$ dipoles and so on. For nuclear structure investigations the most interesting process is the scattering of small dipoles. The smallness of the dipole assures that the interaction with the nucleon is well described by  perturbative QCD.
In the low-$x$ region the small dipoles interact with the nucleon by  gluon exchange only. The color exchange is of short range in contrast to  the interaction of electromagnetic dipoles.
Of particular interest is  the elastic  scattering of color dipoles because the transverse deflection of the dipole measures the spatial distribution of the gluonic matter. 

In this paper we will concentrate on the measurement   of exclusive  $J/\psi$ vector meson photoproduction. The $J/\psi $ meson is a bound state of a charm quark antiquark pair  and therefore the corresponding dipole is naturally small. The cross section is relatively high because we take photoproduction.
 %For $\rho$ and $\phi$ the smallness of the dipole  has to be assured by  a large  photon virtuality,   $Q^2>10$ GeV$^2$, which reduces the cross section substantially. Another advantage of the $J/\psi$ meson is its frequent $\mu^+  \mu^-$ and $e^+ e^-$ decay mode which allows to measure its transverse momentum distribution, $p_{T}$,  very precisely. Also, the narrow width of the $J/\psi$ allows to separate easily the signal from background. 
 %The   $p_{T}$ distribution  of the scattered $J/\psi$ is directly related to the spat%ial distribution of the nuclear matter. 
In nuclei, the small charmed dipole  scatters presumably  on the individual nucleons. In spite of the high energies involved, the nucleus will frequently remain intact because the large absorption cross section together with the optical theorem assures that the scattering process has to be coherent in around 15\% of cases~\cite{KLV,KLMV}.
Since the charmed quark dipole  interacts almost completely via two gluon exchange with  matter the deflection of the $J/\psi$ 
measures directly the {\bf intensity} and the {\bf spatial distribution} of the strong field which keeps the nucleus together. This will allow to see precisely the  structure of gluonic fields  which was never seen before. 
     
%At HERA, the measurement of the $p_T$ distribution of the $J/\psi$ and other vector mesons was only performed on a proton target. The measurements led to an interesting conclusion that 
%the gluonic  radius of the proton, $r_p^{2g}\approx 0.6$ fm, is clearly smaller than the charged radius, $r_p^e=0.875$ and by a factor of almost two smaller than the radius of the nucleon inside the incompressible nuclear matter, $r_N\approx 1.2$ fm.
%These numbers show clearly that much is to be learned by a systematic study of the $p_T$ distribution of $J/\psi$ and other vector mesons together with other DIS reactions.       

The paper is organized as follows; in Section 2 we give an overview of the main properties of the dipole interaction and summarize in some detail the existing experimental results. In Section 3 we discuss the properties of $J/\psi$ as a probe of proton  and nuclei. In Section 4 we discuss the $J/\psi$ production cross section within potential detector acceptances and in  Section 5 we discuss the detector requirements necessary to perform a precise $p_T$ measurement and  we discuss the experimental requirements necessary to assure the elastic signature. In Section 6 we summarize the results and in 
 Section 7 we conclude.  In  the Appendix we give a derivation of the dipole representation in a simple case of elastic photon-proton scattering.

\section{Dipole description of DIS reactions}
Dipole interactions are, in a way, as fundamental as  Rutherford scattering. In  Rutherford scattering the incoming electron emits a virtual photon which scans the charge distribution of the nuclear target. In elastic dipole scattering the incoming electron emits a virtual photon which turns into a small quark antiquark pair. In  leading order  QCD, at large virtuality scales and high energies, the $q\bar q$ pair  interacts elastically with the nucleon by exchanging {\it two} gluons with large transverse momenta, $\vec{l}$ and $\vec{l} +\vec{\Delta}$. The transverse momenta of the gluons cannot be directly observed, however  their difference, $\vec{\Delta}$, is measurable in  elastic vector meson scattering
because it is equal  
 to the difference between  the transverse momentum of the incoming virtual photon and the final vector meson.     The two gluon interaction leads in QCD (together with QED) to the dipole representation. In the Appendix we derive  the dipole representation from the Feynman diagrams in a simple case of virtual photon-proton elastic scattering, when $\vec{\Delta}=0$. In the following we discuss the main properties of the dipole representation. 
\begin{figure}[htbp] %  figure placement: here, top, bottom, or page
   \centering
   \includegraphics[width=4.2in]{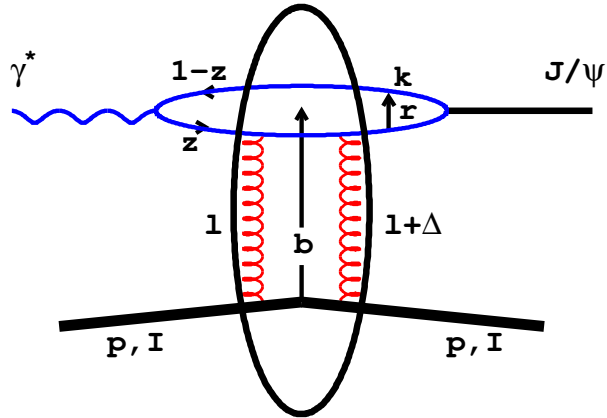} 
   \caption{Elastic  scattering of a $J/\psi$ meson  on a proton or ion
 in the dipole representation.}
   \label{fig:quasi-elastic}
\end{figure}

%In the proton rest frame, the dipole life-time is much longer than the life-time of its interaction with the target proton.

\subsection{Dipole representation}
   
The dipole interaction 
proceeds in three stages: first the incoming virtual photon fluctuates into a quark-antiquark pair, then the $q\bar{q}$ pair elastically scatters on the target, and finally the  $q\bar{q}$ pair recombines to form a final state. 
The creation, scattering and recombination of the dipole occur  in different space time regions because 
the life time $\tau$ of the $q\bar q$ fluctuation is very long. In the proton rest frame,
$\tau \approx 1/m_px$. 
Even at $x=10^{-2}$ this correspond to the distance of $\sim 20$ fm, which is larger than the size of all nuclei, $R\approx 1.2\cdot A^{1/3}$ fm.
 In addition, due to high energies,  the transverse positions of the quark and antiquark do not change in the scattering process  so that the dipole is not changing its size (see Appendix).  

The amplitude for the   elastic photon-proton scattering,  $\gamma^* p\rightarrow \gamma^* p$,
  $\mathcal{A}^{\gamma^*p}(x,Q,\Delta)$, is therefore the product of amplitudes of these three subprocesses integrated over the dipole variables $\vec{r}$ and $z$:
\begin{eqnarray}
  \mathcal{A}^{\gamma^*p}(x,Q,\Delta) = \sum_f  \int\!\dif^2\vec{r}\,\int_0^1\!\frac{\dif{z}}{4\pi}\,\Psi^*(r,z,Q)\,\mathcal{A}_{q\bar q}(x,r,\Delta)\,\Psi(r,z,Q),
  \label{eq:elamp}
\end{eqnarray}
where $\Psi(r,z,Q)$ denotes the amplitude for the incoming virtual photon, with virtuality $Q$, to fluctuate into a quark--antiquark dipole with flavour $f$\footnote{We suppress here references to the photon helicities for simplicity. }. The wave function is determined 
from light cone perturbation theory to leading order in the fermionic charge, see Appendix. 
The amplitude for the $q \bar q$ to recombine to a virtual photon is $\Psi^*(r,z,Q)$.
 $\mathcal{A}_{q\bar q}(x,r,\Delta)$ is the elementary amplitude for the scattering of a dipole of size $r$ on the proton, $\vec{\Delta}$ denotes the transverse momentum transfered from the dipole to  the  target, and $x$ is the Bjorken variable. The sum  should be taken over all quarks flavours $f$, including charm. For charmed quark the definition of the Bjorken $x=x_B$ should be replaced by $x=x_B(1+4m_{ch}^2/Q^2)$ to take into account the charm threshold effects.  
 For similar reasons, in the  case of vector meson production, $x=x_B(1+M_V^2/Q^2)$.

The elementary elastic amplitude $\mathcal{A}_{q\bar q}$ is defined such that the elastic differential cross section for the $q\bar{q}$ pair scattering on the proton is
\begin{eqnarray}
  \frac{\dif\sigma_{q\bar q}}{\dif t} = \frac{1}{16\pi}\left\lvert\mathcal{A}_{q\bar q}(x,r,\Delta)\right\rvert^2,
  \label{eq:elsiggp}
\end{eqnarray}
where $t=-\vec{\Delta}^2$. The notation follows  the conventions of ref.~\cite{MSM,KMW}.
Elastic dipole scattering is connected by the  optical theorem to  the total cross section for photon-proton or photon-nucleus scattering. 
%The optical theorem assures also the connection to various diffractive reactions like the inclusive diffraction, DVCS process and diffractive jet production. By assuming a vector meson wave function it is also
%directly connected  to exclusive diffractive vector-meson processes.
To evaluate the connections between the total cross section and various diffractive reactions it is convenient to work in coordinate space and define the S-matrix element at a particular impact parameter~$b$
\begin{eqnarray}
  \mathcal{A}_{q\bar q}(x,r,\Delta) = \int\!\dif^2\vec{b}\;\mathrm{e}^{-\mathrm{i}\vec{b}\cdot\vec{\Delta}}\,\mathcal{A}_{q\bar q}(x,r,b)
  = \mathrm{i}\,\int \dif^2\vec{b}\;\mathrm{e}^{-\mathrm{i}\vec{b}\cdot\vec{\Delta}}\,2\left[1-S(x,r,b)\right]. 
  \label{eq:smatrix}
\end{eqnarray}
This corresponds to the intuitive notion of impact parameter when the dipole size is small compared to the size of the proton.  The optical theorem then connects the total cross section for the $q\bar q$ pair scattering on the proton to the imaginary part of the forward scattering amplitude:
\begin{eqnarray}
  \sigma_{q\bar q}(x,r)=\mathrm{Im}\,\mathcal{A}_{q\bar q}(x,r,\Delta=0)=\int \dif^2 \vec{b}\; 2[1-\mathrm{Re}\,S(x,r,b)].
  \label{eq:totsiggp}
\end{eqnarray}
The integration over $\vec{b}$ of the S-matrix element motivates the definition of the differential dipole cross section as
\begin{eqnarray}
  \frac{\dif\sigma_{q\bar q}}{\dif^2 \vec{b}}= 2[1-\mathrm{Re}\,S(x,r,b)].
  \label{eq:difsiggp}
\end{eqnarray}

The total cross section for $\gamma^* p$ scattering, or equivalently $F_2$, is obtained, using \eqref{eq:elamp} and \eqref{eq:totsiggp}, by integrating the dipole cross section with the photon wave functions: 
\begin{eqnarray}
  \sigma^{\gamma^* p}_{T,L}(x,Q) 
  =  \mathrm{Im}\,\mathcal{A}^{\gamma^* p}_{T,L}(x,Q,\Delta=0) 
  = \sum_f \int\!\dif^2\vec{r} \int_0^1\!\frac{\dif z}{4\pi}
  (\Psi^{*}\Psi)_{T,L}^f
  \, \sigma_{q\bar q}(x,r),
  \label{eq:siggp}
\end{eqnarray}
where  $_T$ or $_L$ denotes the transverse or longitudinal polarization of the incoming photon.
For completeness we also give  the relation between   the $F_2$ structure function and the $\gamma^* p$ cross section at small-$x$, 
$$
F_2(x,Q^2) = \frac{Q^2}{4\pi^2 \alpha_{em}} (\sigma_T+\sigma_L) .
$$

Elastic vector-meson production appears  in a similarly transparent way. The amplitude is given by
\begin{eqnarray}
A_{\gamma^* p\rightarrow pV}(\Delta) = \int d^2r \int \frac{dz}{4\pi} \int d^2b\, \Psi^*_V\Psi \exp(-i\vec{b}\cdot\vec{\Delta}) 2[1- S(b)].
\label{eq:amvecm}
\end{eqnarray}
 Assuming that the S-matrix element is predominantly real, we may substitute $2[1-  S(b)]$ with $d \sigma_{q\bar q}/d^2b$. Then, the elastic cross section is 
\begin{eqnarray}
\frac{d\sigma^{\gamma^* p\rightarrow Vp}}{dt} = 
\frac{1}{16\pi}\left|\int d^2r \int \frac{dz}{4\pi} \int d^2b \, \Psi^*_V\Psi \exp(-i\vec{b}\cdot\vec{\Delta}) \frac{d\sigma_{q\bar q}}{d^2b}\right|^2.
\label{eq:xsvecm}
\end{eqnarray}
The equations (\ref{eq:siggp}) and (\ref{eq:xsvecm}) determine the total DIS cross sections and the exclusive diffractive vector-meson production cross section. 
Fig.~\ref{fig:quasi-elastic} shows  a diagram of the dipole scattering when the final state 
is a $J/\psi$ meson.

\subsection{Dipole cross sections}

\begin{figure}[htp]
\vspace{0.cm}
\hspace{0.cm}
   \centering
 \includegraphics[width=0.7\textwidth]{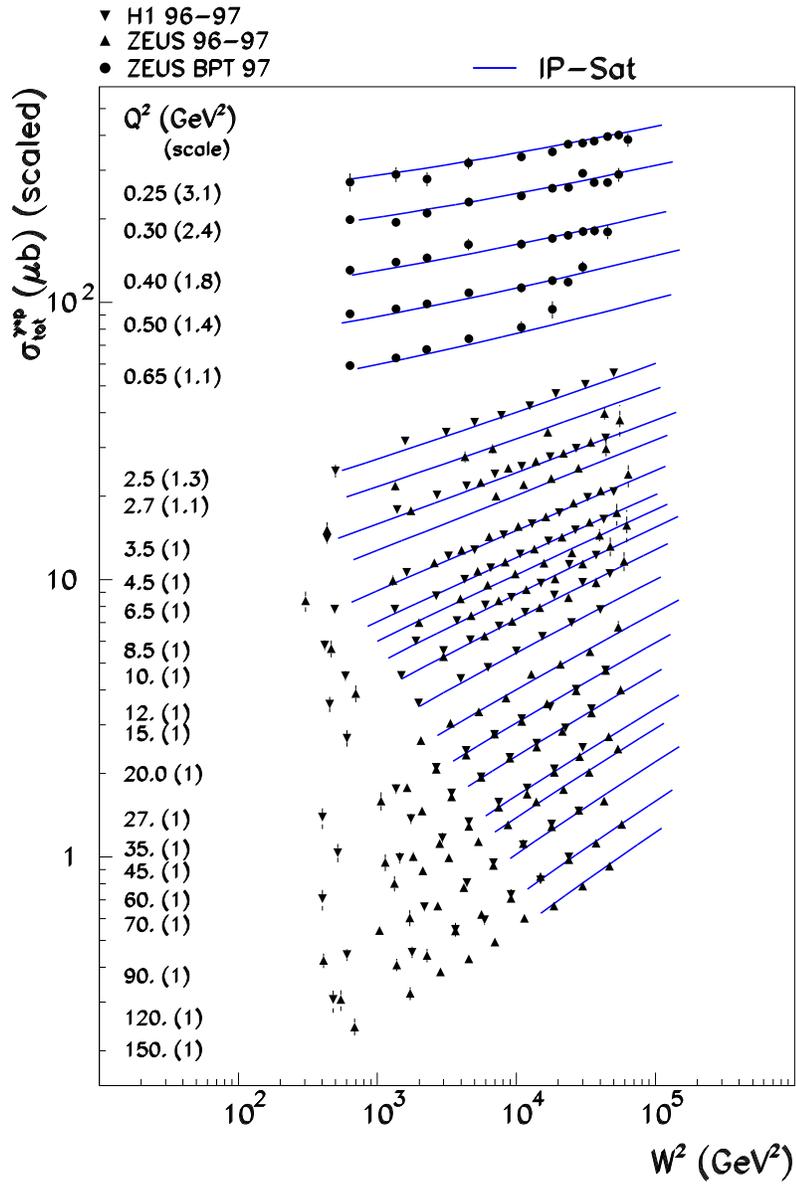}
\vspace{ 0.cm} 
\caption{ The $\gamma^* p$ cross section as a function of $W^2$. The solid lines are showing the dipole fit to the data~\cite{KT} in the low-$x$ region, 
$x<10^{-2}$.} 
\label{fig:sigtot}
\end{figure}

The universal dipole cross section $\dif\sigma_{q\bar q}/\dif^2 \vec{b}$  contains all  the  information about the gluon 
content of the target and the QCD evolution of the gluon density.
Its particular form depends on the evolution schema. In the  DGLAP formalism,  the dipole cross section for small dipoles is given by\footnote{The derivation of the dipole cross section   is given in the Appendix}, 
\begin{eqnarray}
\frac{d\sigma_{q\bar q}}{d^2b} = \frac{\pi^2}{3}\, r^2\, \alpha_s(\mu^2)\, xg(x,\mu^2)\, T(b).
\label{eq:dxs}
\end{eqnarray}
 Here
 $xg(x,\mu^2)$ is the gluon density and $\mu^2$ denotes the evolution scale. 
 The starting scale is denoted by $\mu_0^2$. 
The scale is given by the inverse of the dipole size and is usually taken as
 $\mu^2 = 4/r^2 + \mu_0^2$, to stabilize the $\alpha_s$ behavior for large dipoles.  The function
 $T(b)$ is the transverse profile of the gluon density of the proton. 
 It is convenient to write the gluon density in a semi-factorized form, $xg(x,\mu^2)T(b,x)$. Because the $x$ dependence of $T(b)$ is weak it is omitted in the following discussion.
  At every $x$  the transverse profile is normalized to 1,  
 $\int d^2b T(b) = 1$.     
The parameters of the gluon density  
are determined from the fit to the total inclusive DIS cross section using eq.~\ref{eq:siggp}. The fit is shown in Fig.~\ref{fig:sigtot}.
  The gluon density at a scale $\mu_0^2$ is parametrized by $xg(x,\mu_0^2)=c\cdot x^{-\lambda}$. The three free parameters   are  $c$, $\lambda$ and $\mu_0^2$. The prediction of the model, at low $Q^2$ values,  also depend on the  assumption on the quark masses, $m_f$.  Other schemas like the  CGC evolution or Regge parametrizations lead also to a successful description of the data ~\cite{IIM,WK,Forshaw2006}. More details can be found in~\cite{KMW,WK}.

\begin{figure}
  \centering
  \includegraphics[width=0.65\textwidth]{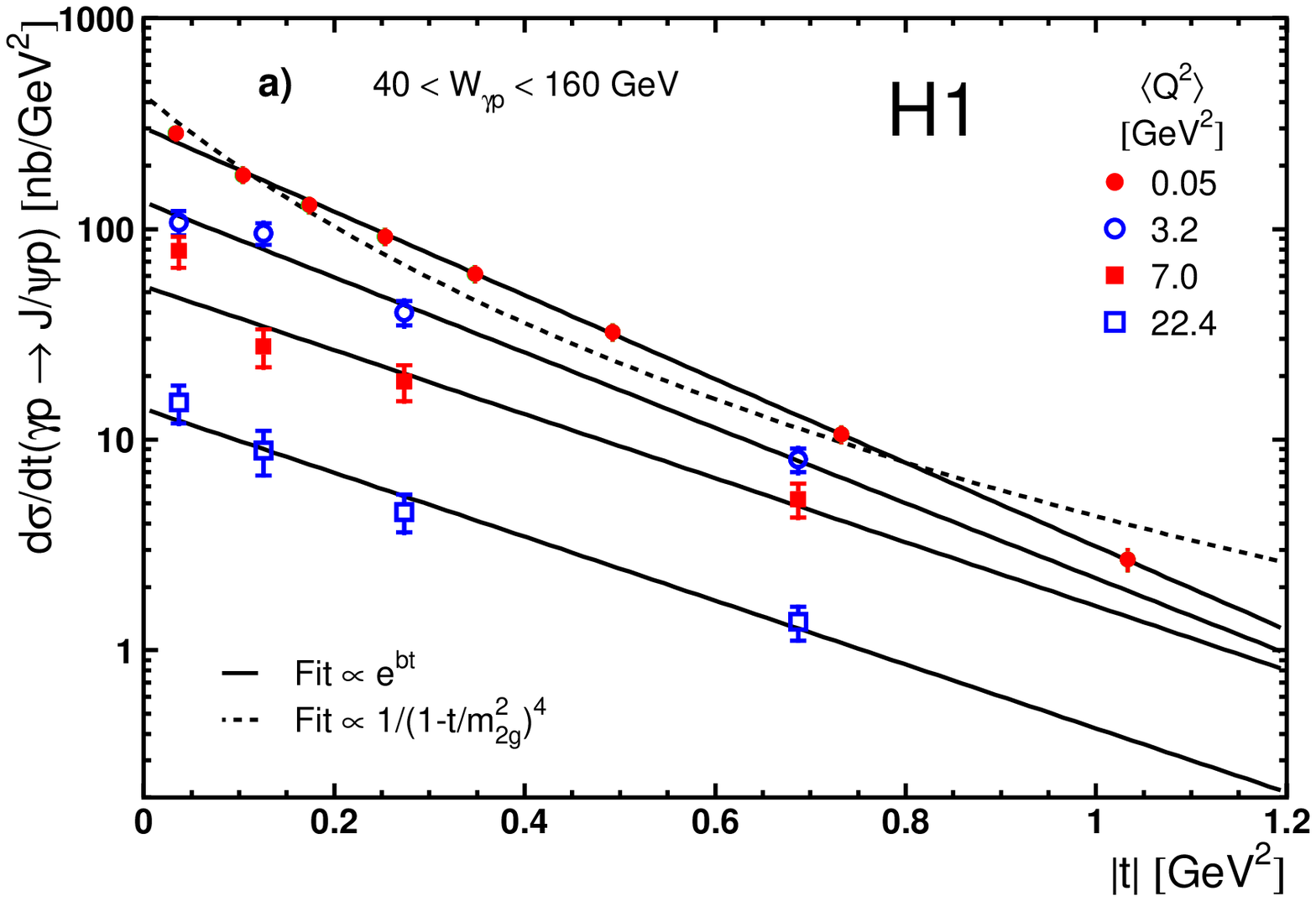}
  \caption{Differential cross section for the elastic $J/\psi$  production~\cite{H1jpsi}. }
  \label{fig:t_jpsi}
\end{figure}

The transverse profile is assumed to have a two dimensional Gaussian form:
\begin{eqnarray}
T(b) = \frac{1}{2\pi B_G}\exp(-b^2/2B_G),
\label{eq:Tofb}
\end{eqnarray}
since  the measured $t$-distributions were found to be well described  by the exponentially falling distribution,
$ d\sigma_{VM}^{\gamma^{*}p}/dt \propto \exp(-B_D |t|)$,  see Fig.~\ref{fig:t_jpsi}\footnote{
The Fourier transform of a gaussian in $\vec{b}$ is a gaussian in $\vec{\Delta}$ corresponding to an exponential in $t$. }.  The coefficient $B_D$ is not exactly equal to $B_G$ because the size of the vector meson is not negligible (see the discussion in Section 2.4). In addition,
 the observed coefficient $B_D$ shows some $x$ dependence (e.g. in $J/\psi$ photoproduction) and so   $B_G$ should be  $x$ dependent (see the discussion in Section 3.1).
 We note also that the fit with the  dipole form factor,
$d\sigma/dt \propto 1/(1-t/M^2)^4$, which is usually used to parametrize the proton form factors does not describe data appropriately~\cite{H1jpsi}.

The dipole cross section, eq.~\ref{eq:dxs}, is propotional to the dipole area, $ r^2$, and the gluon density seen by the dipole, $xg(x,\mu^2)$. It is therefore a measure of the gluon density. When either the dipole size or the gluon density is not small it is  convenient to use 
the eikonalized form of the dipole cross section to take into account possible saturation effects, 
\begin{eqnarray}
\frac{d\sigma_{qq}}{d^2b} = 2\, \left(1-\exp(-\frac{\pi^2 r^2\, \alpha_s(\mu^2)\, xg(x,\mu^2)\, T(b )}{2\cdot 3})\right )=2\, \left(1-\exp(-\frac{\Omega }{2})\right). 
\label{eq:xdip}
\end{eqnarray}
Here $\Omega$ denotes the opacity which is equal to the right side of 
eq.~\ref{eq:dxs}.
The above formula is called the Glauber-Mueller dipole cross section. A diffractive cross section of this type was used around 50 years ago to study the diffractive dissociation of  deuterons by Glauber~\cite{Glauber:1955} and reintroduced by Mueller~\cite{Mueller:1989} to describe  dipole scattering in deep inelastic processes.  In contrast to  Glauber scattering, in DIS the functional form of the opacity is known from QCD.

%All these processes were measured at HERA and dipole predictions are in good agreement
%  with the data. This means that  higher order QCD corrections are small and th%at  dipole scattering is a fundamental QCD process. 

\subsection{DVCS and VM production}

\begin{figure}
  \centering
  \includegraphics[width=0.4\textwidth]{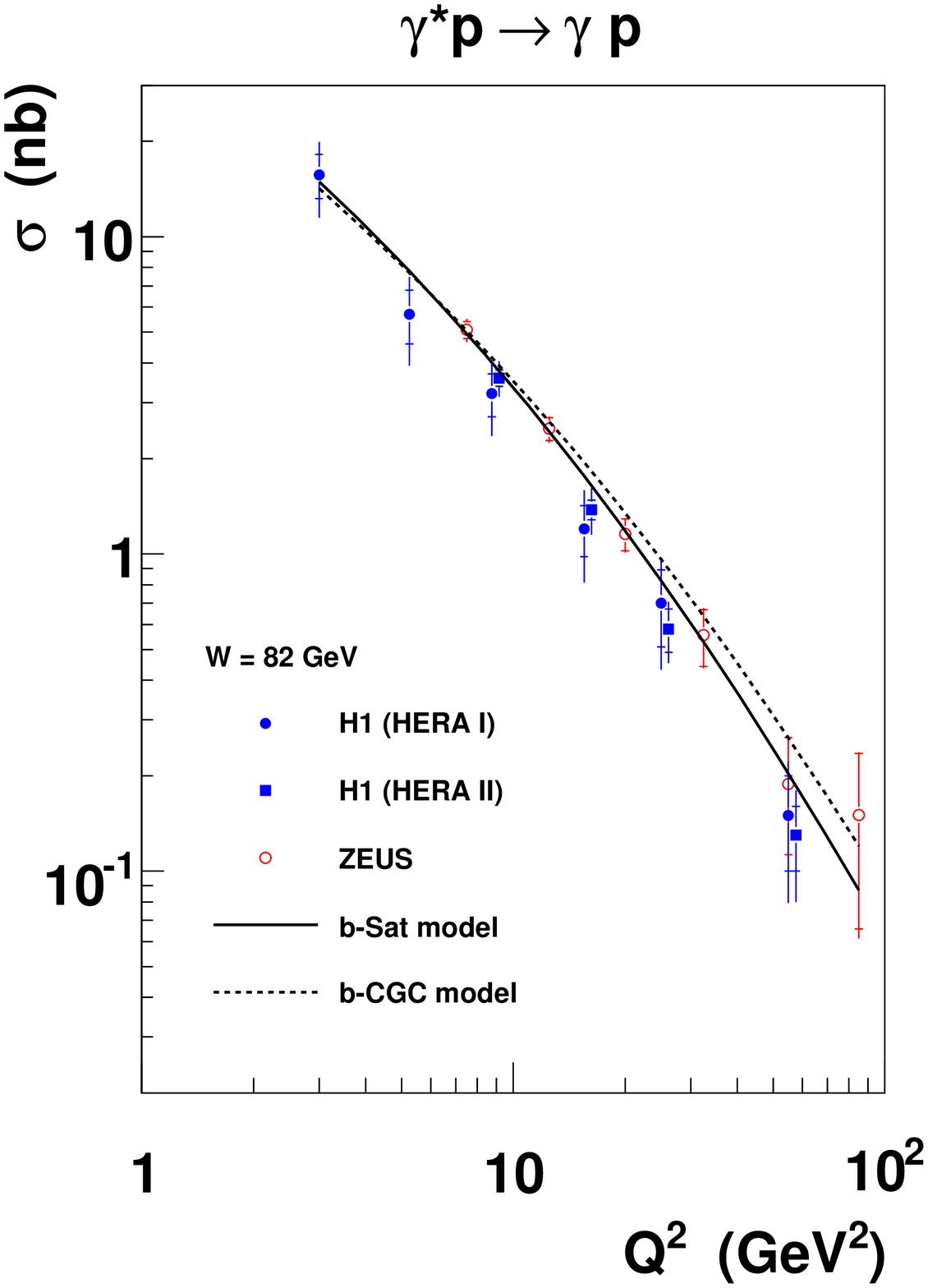}
  \includegraphics[width=0.4\textwidth]{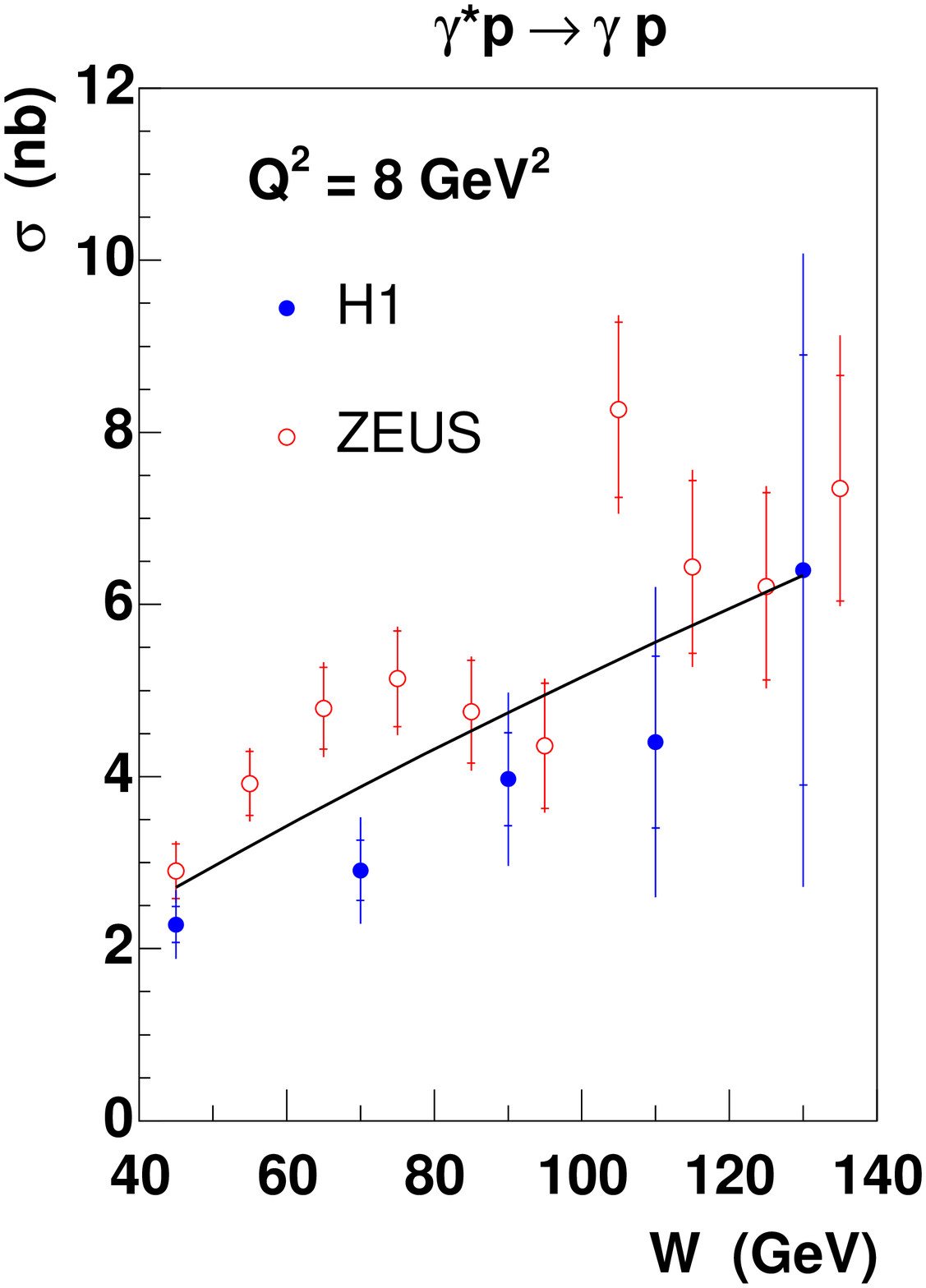}
  \caption{The total DVCS cross sections $\sigma$ as a function of~$Q^2$ (left) and $W$ (right)~\cite{DVCSdata} compared to predictions from the b-Sat and b-CGC models~\cite{WK}.}
  \label{fig:crossq_dvcs}
\end{figure}

The elastic  cross section, eq.~\ref{eq:xsvecm}, was derived under the assumption that the dipole size is much smaller than the proton size. For dipoles 
with a size $r$ the explicit  QCD calculation~\cite{Bartels:2003yj} shows that
$\vec{\Delta}$ conjugates to $\vec{b}+(-z)\vec{r}$. Therefore, the cross section is not only sensitive to the  proton impact factor, $b$, but also to the dipole size, $r$.
The modified dipole cross section is then
\begin{eqnarray}
\frac{d\sigma^{\gamma^* p\rightarrow Vp}}{dt} = 
\frac{1}{16\pi}\left|\int d^2r \int \frac{dz}{4\pi} \int d^2b \, \Psi^*_V\Psi \exp(-i[\vec{b}+(1-z)\vec{r}]\cdot\vec{\Delta}) \frac{d\sigma_{q\bar q}}{d^2b}\right|^2.
\label{eq:xsvecma}
\end{eqnarray}
This  cross section can   also be used  to describe the DVCS process for which the final state consists of the scattered electron, proton and a real photon. The wave function for the outgoing state is just the amplitude $\Psi^*(Q^2=0)$ for the real photons. 
Equation~\ref{eq:xsvecma} then gives, after small correction for the real part\footnote{The derivation of the cross section for  exclusive vector meson production or DVCS  relies on the assumption that the scattering amplitude is purely imaginary. The real part of the amplitude can be accounted for by multiplying the exclusive cross section by a small correction factor  given in ref.~\cite{Forshaw2003}.} 
 and skewedness\footnote{For vector meson production or DVCS one should use the off-diagonal gluon densities, since here the two gluons carry different fractions $x$ and $x'$ of the proton momentum. This effect can be taken into account by multiplying the gluon distribution $xg(x,\mu^2)$ by a correction factor, $R_g$ given in~\cite{Shuvaev:1999ce}.}
   are applied, an absolute prediction for the DVCS process. The prediction is shown along with HERA data in Fig~\ref{fig:crossq_dvcs}. The figure shows an impressive agreement between the dipole model predictions and the data if we realize that  two very different processes are described with the same amplitudes; the average event which contributes to the total cross section has around 40 particles in the final state, while the DVCS event has  just a proton, a photon and an electron. 
The wave function of the final state, $\Psi^*(Q^2=0)$, is very different than that of the incoming virtual photon. And yet   
 all the distributions and the absolute event rates of the DVCS process are properly described. 
%Although the precision of each data point is only on the level of 10\% the agreement of so many measurements indicate an agreement on the level of better than 1\% (Allen is it right?). 
This means that the net effects of  possible higher order QCD radiative corrections~\cite{KlKr} or  possible kinematical corrections~\cite{ref.KMS}, have to be  small or are consistently absorbed into the gluon density. 

\begin{figure}
  \centering
  \includegraphics[width=0.33\textwidth]{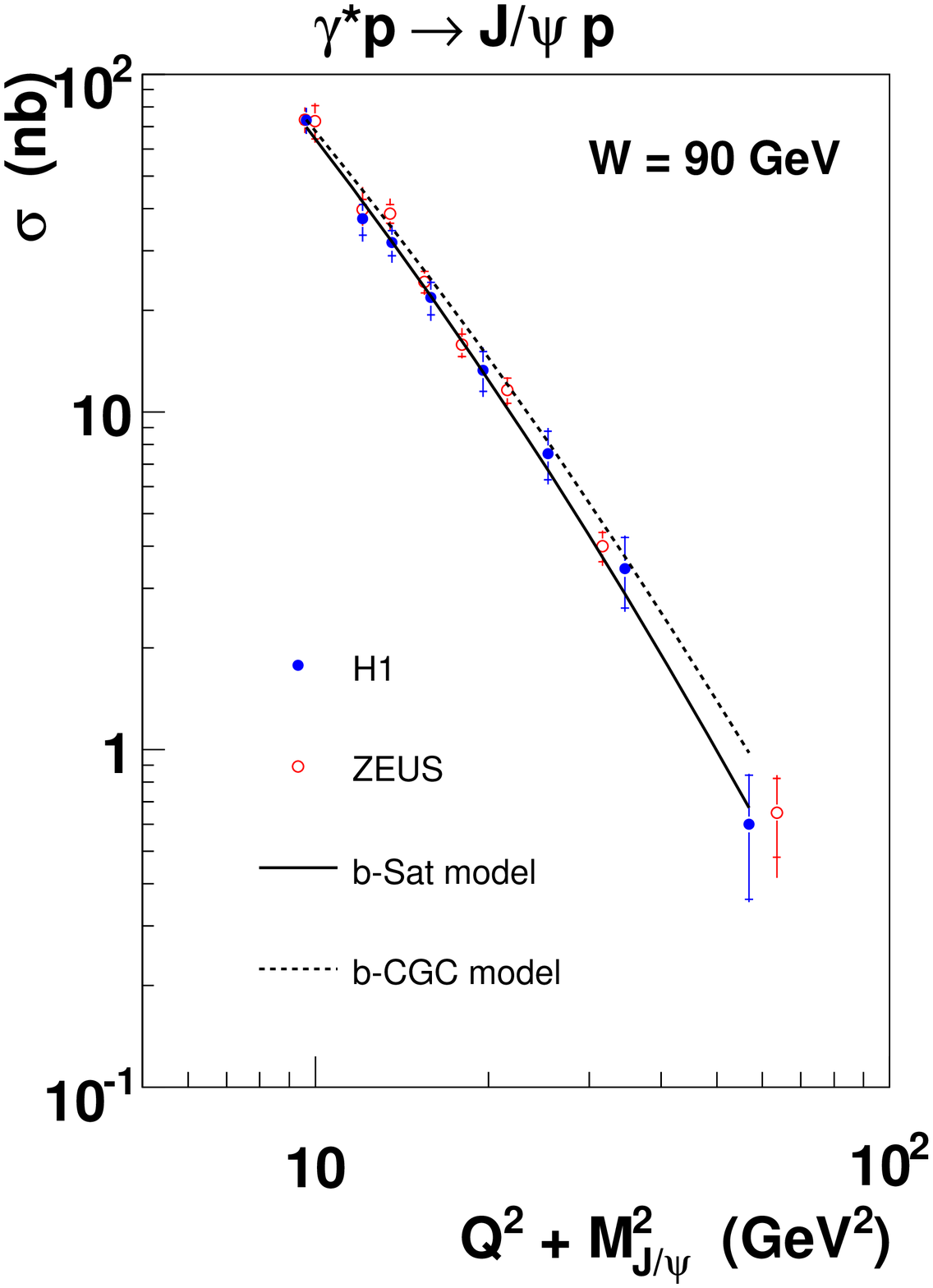}%
  \includegraphics[width=0.33\textwidth]{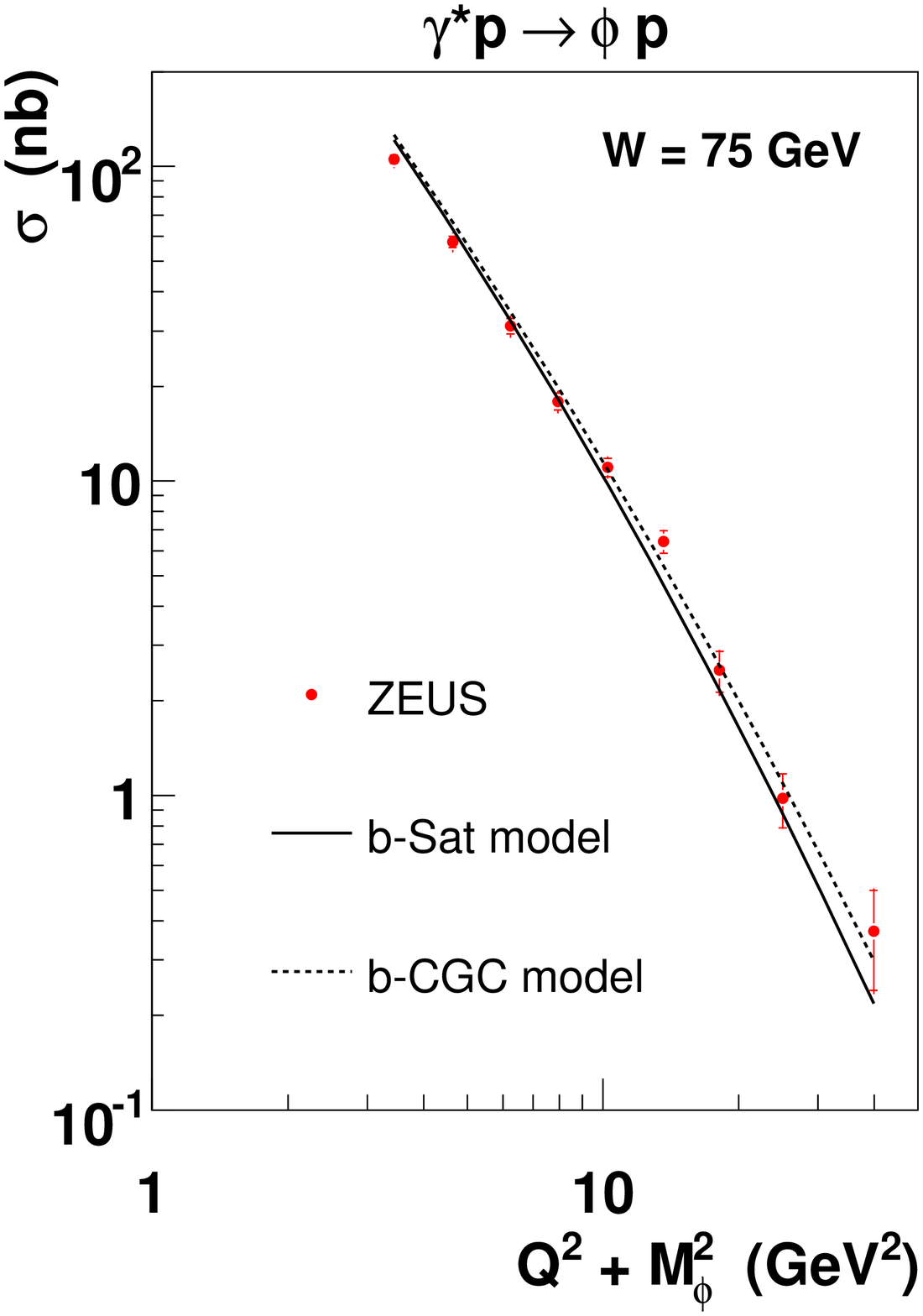}%
  \includegraphics[width=0.33\textwidth]{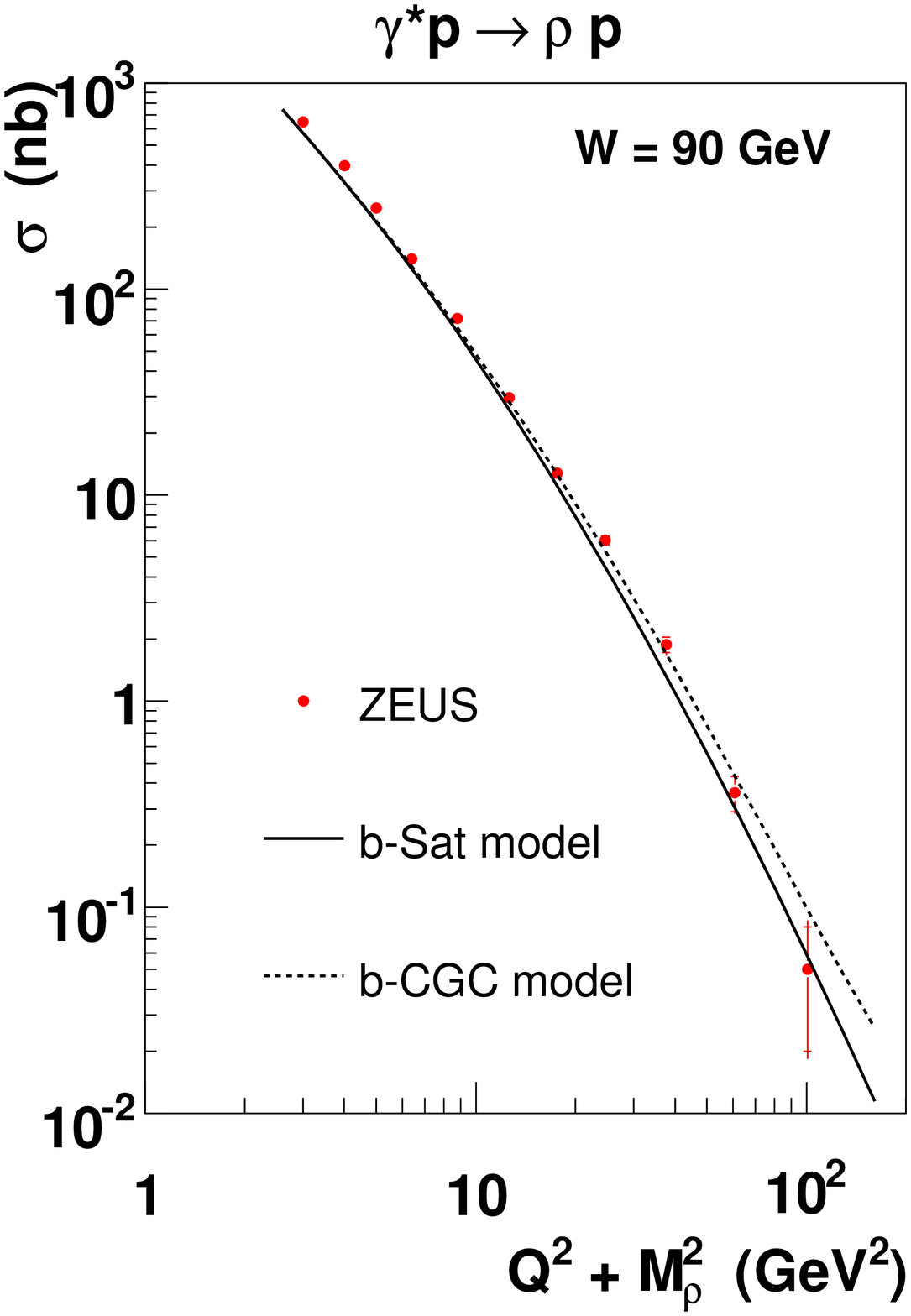}
  \caption{The total cross section $\sigma$ vs.~$(Q^2+M_V^2)$ for exclusive $J/\psi$ \cite{H1jpsi,ZEUSjpsi}, $\phi$ \cite{ZEUSphi} and $\rho$ \cite{ZEUSrho} meson production compared to predictions from the b-Sat and b-CGC models using the ``boosted Gaussian'' vector meson wave function~\cite{WK}.}
  \label{fig:crossq}
\end{figure}

Successful comparisons of the inclusive diffractive cross sections with the dipole model predictions were performed in several investigations~\cite{GBW,BGK,Forshaw2006,KLMV}.   
Inclusive diffraction denotes the sum of all processes in which the proton remains intact.  Since the summation is over a complete set of states the cross section depends only on the  photon wave functions. However, the dipole cross section in this case has to take into account also the possible $q\bar q g$ Fock state which makes the comparison less clean than in the DVCS case. Nevertheless, the comparison of the cross sections with the predictions works again very well~\cite{KLMV}. The same is also true for the comparison with the diffractive jet cross sections~\cite{ZEUSdiffjets}.

Further support for the validity of the dipole picture comes from the very good agreement of the predicted vector meson cross sections with the data, shown in Fig.~\ref{fig:crossq}. The vector meson wave functions are constructed according to a standard procedure developed 
in~\cite{Nemchik:1994fp,Nemchik:1996cw}. They are educated guesses based on  general considerations described in~\cite{Forshaw2003}, ~\cite{KMW}. The only phenomenological input which is necessary  is obtained from the measured electronic decay width of the vector meson.   
The very good agreement between data and dipole model predictions means that the vector meson wave functions  were  properly estimated. This is further supported by the excellent description of $\sigma_L/\sigma_T$ ratio as a function of $Q^2$ for the $J/\psi$ and $\phi$ vector mesons~\cite{KMW}.
In case of $\rho $ mesons this ratio is not so well described. This can be attributed to the complicated  dynamic of the $\rho$  decay into  pions. 

\subsection{The $t$-distributions}

At HERA, the distributions of the square of the  momentum transfer, $t=-\vec{\Delta}^2$, have been  measured  in  exclusive vector meson and DVCS processes. The value of $t$ is usually  determined from the $p_T$ imbalance of the final state particles seen in the detector.  The $t$ distributions have also been measured  using  forward proton spectrometers  for all diffractive processes.  The forward spectrometers have however   a relatively small acceptance. 

All measured   $t$ distributions are very well described by the exponential dependence
$d\sigma/dt\propto \exp(-B_D|t|)$. 
The $t$ range used in the fit to the vector meson differential cross sections is typically  limited to $|t|< 1$ GeV$^2$  to avoid regions where large corrections were applied to account for  the proton dissociation process. 
$B_D$ is related to the size of the interaction region, as discussed below.
It is sensitive to the gluonic proton shape and to the sizes of the dipoles  contributing to the process.
The dipole sizes depend, through the incoming and outgoing wave functions, on the $Q^2$ value and the vector meson spatial extension.   
%A similar form of the impact parameter dependence  is also  derived from the generalized parton densities,~\cite{Diehl,FrST}.

 \begin{figure}
  \centering
  \includegraphics[width=0.7\textwidth]{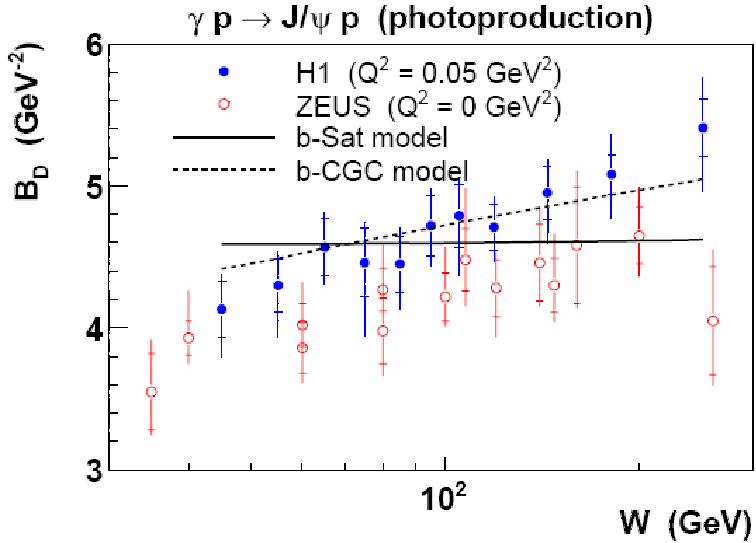}
  \caption{The $t$ slope parameter $B_D$ vs $W$ for  $J/\psi$ photoproduction.}
  \label{fig:bd_jpsi}
\end{figure}
\begin{figure}
  \centering
  \includegraphics[width=0.33\textwidth]{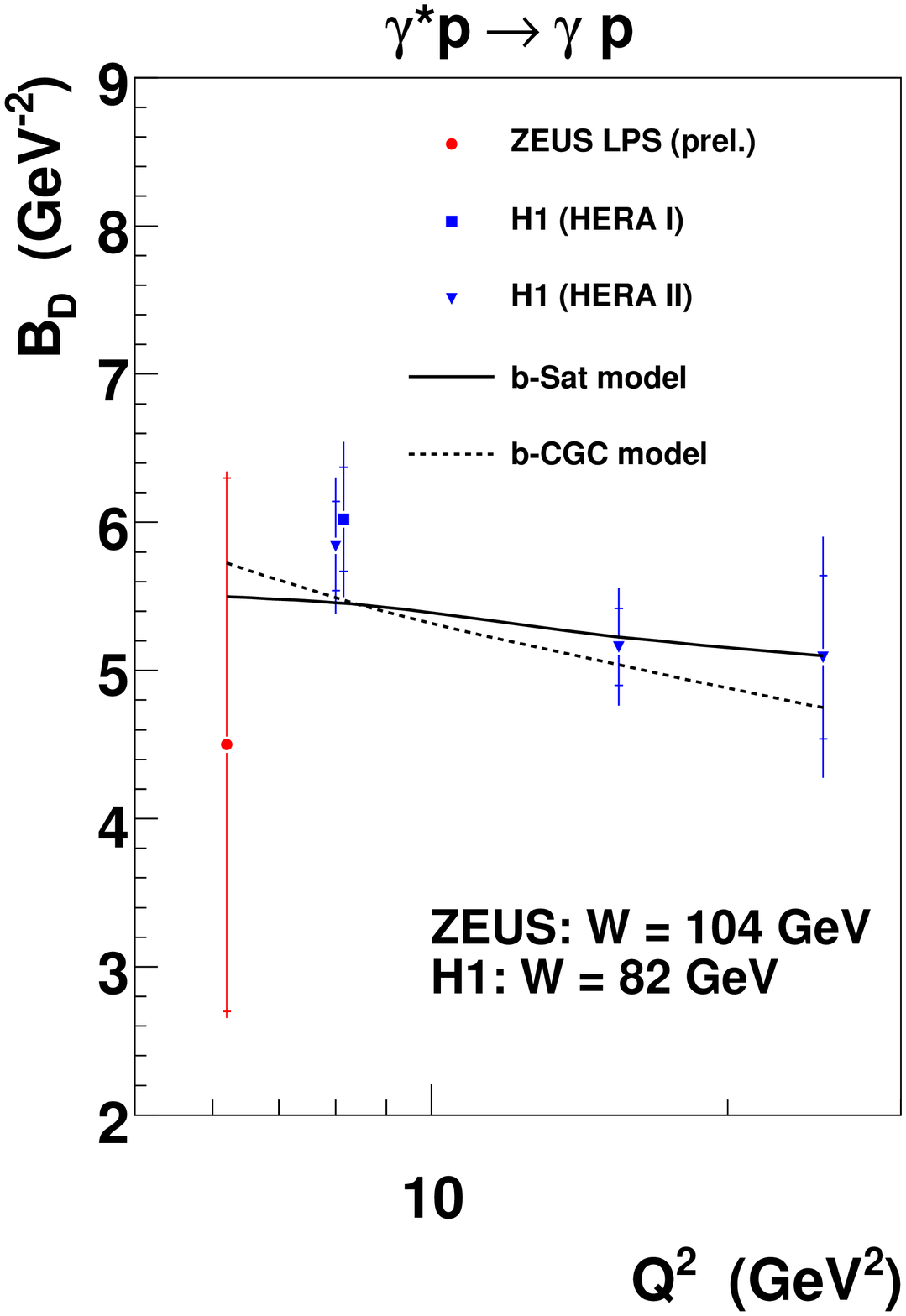}%
  \includegraphics[width=0.33\textwidth]{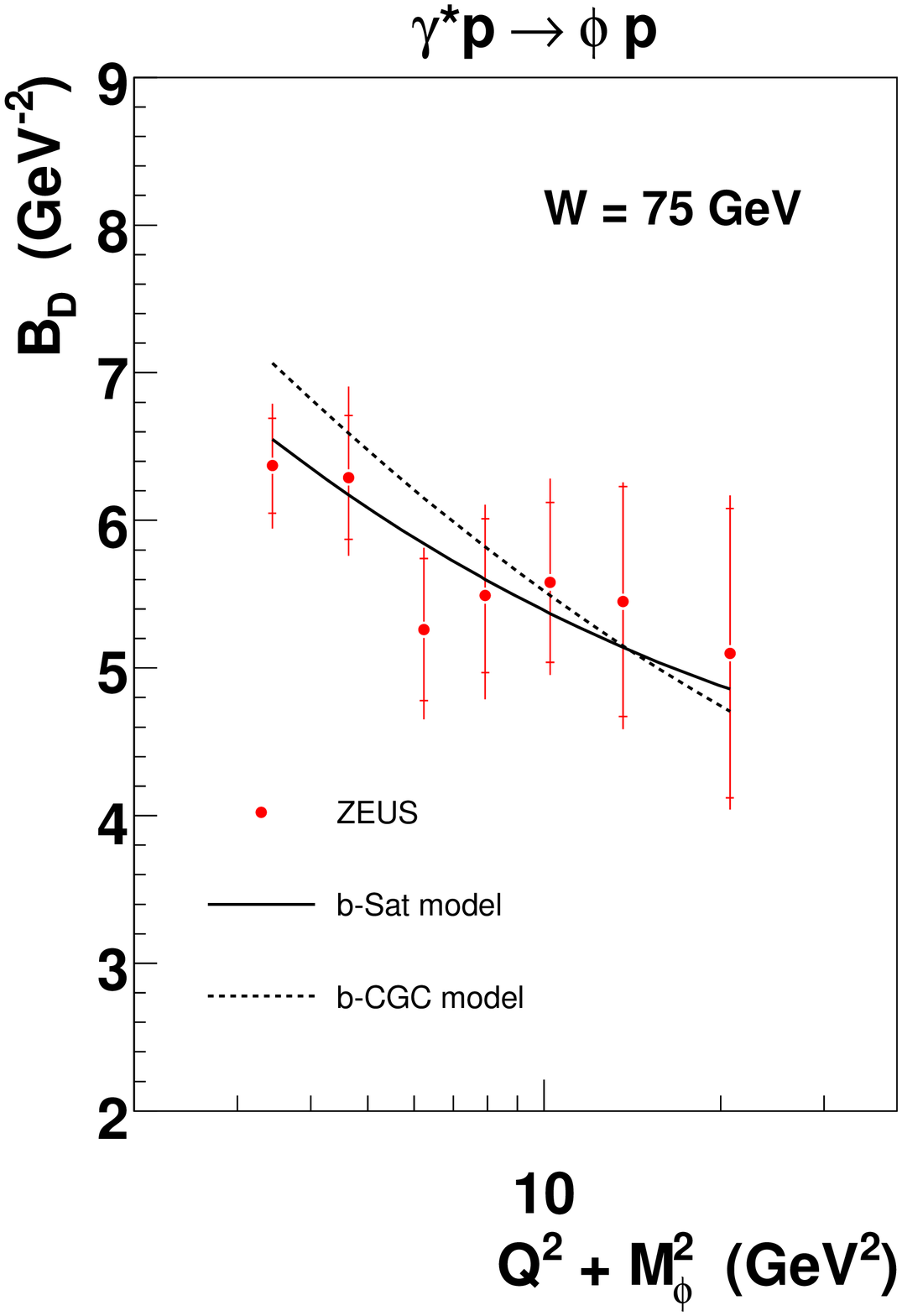}%
  \includegraphics[width=0.33\textwidth]{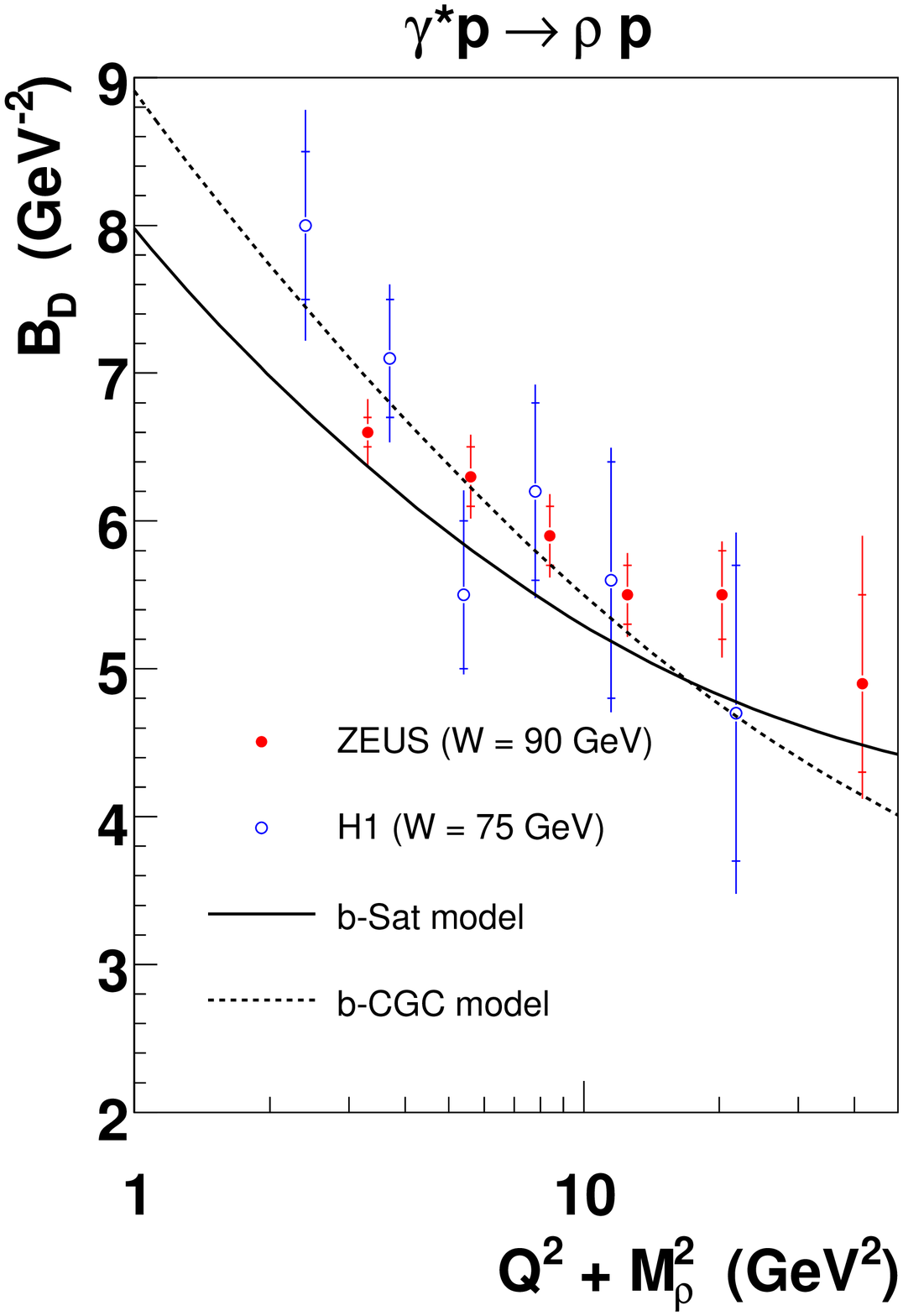}%
  \caption{The $t$-slope parameter $B_D$ for the DVCS process and for exclusive  $\phi$ \cite{ZEUSphi} and $\rho$ \cite{Adloff:1999kg,ZEUSrho} meson production compared to predictions from the b-Sat and b-CGC 
models.}
  \label{fig:bd}
\end{figure}

  Fig.~\ref{fig:bd_jpsi} shows  $B_D$ as a function of $W$ for $J/\psi$ photoproduction. The average value of $B_D$ from this process  was used to fix the $B_G$ parameter of the proton shape in the b-Sat model to
$B_G=4.0$ GeV$^{-2}$.
 The model allows then to predict the expected $B_D$'s for other processes, like DVCS and  $\phi$ and $\rho$ production. The predictions are compared to data in Fig.~\ref{fig:bd}. The value of $B_D$ depends considerably on $Q^2$ and the type of process  
since the effective size of vector mesons and of the DVCS  photon differs substantially. The dipole model reproduces the changes in the value of $B_D$  very well. 

As seen in Fig.~\ref{fig:bd_jpsi} 
the value of $B_D$ in  $J/\psi$ photoproduction  shows some $W$ dependence. This can be attributed to QCD evolution effects. 
In the DGLAP evolution scheme, due to strong ordering, the increase of $B_D$ with decreasing $x\approx M_{J/\psi}^2/W^2$ is expected to be negligible~\cite{BartKow}, as in the b-Sat model.
In the BFKL evolution scheme the evolution in $x$  could lead to a small increase of the size of the interaction area. This was taken into account in the b-CGC model, which is based on  a phenomenological approach to  the BFKL and BK equations~\cite{IIM,KMW,WK}. Alternatively, the variation  of $B_D$ with $W$ can also be attributed to the increase of the contribution of the "pion cloud" at small $x$~\cite{StrWeis}.

\section{$J/\psi$ as a probe of  proton and nuclei}

The properties of diffractive processes described in the previous section single out  $J/\psi$ photoproduction as an ideal probe for investigation of nuclear properties due to three main reasons:
\begin{itemize}
\item
The $J/\psi$  meson provides the  smallest probe when compared to the other diffractive processes measured at HERA. This can be directly seen from the measurement of the sizes of the interaction region, $B_D$, shown in Fig.~\ref{fig:bd_jpsi} and~\ref{fig:bd}.       
\item
The observed number of well measured events is substantially larger for
$J/\psi$ photoproduction than for other exclusive diffractive processes with  similar probe size.\footnote{ 
This is due to the fact that the cross section for the electroproducion process is approximately proportional to $\log(Q^2_{max}/Q^2_{min})$ 
and to the photoproduction cross section $\sigma^{\gamma p}$.
In the ZEUS and H1 measurements of  $J/\psi$ photoproduction,  
$Q^2_{min}\approx 10^{-12}$ GeV$^2$ and $Q^2_{max}\approx 1$ GeV$^2$. In this Q$^2$ region the photoproduction cross section, $\sigma^{\gamma p}$, is almost constant. For the $\rho$, $\phi$ or DVCS process one has to require that $Q^2_{min}\approx 10$ GeV$^2$ to assure that the interaction is mediated by the small dipoles. In addition, the virtual photon-proton cross section is droping quickly with increasing $Q^2$, $\sigma^{\gamma^* p} \sim 1/(M_V^2+Q^2)^n$, with $n
\approx 3$, which limits effectively the Q$^2$ range.} 
In the central region of the H1 and ZEUS detectors, the number of well measured $\gamma p\rightarrow J/\psi\, p\rightarrow \mu\mu\, p$ and $\gamma p\rightarrow J/\psi\, p\rightarrow ee\,p$ events is about 10 times higher than for $\gamma^* p\rightarrow \rho p\rightarrow \pi\pi\, p$ and by a factor of 40 higher than for $\gamma^* p\rightarrow \phi p\rightarrow KK\, p$ processes with $Q^2+M_V^2>10$ GeV$^2$. For the DVCS reaction, with $Q^2>10$ GeV$^2$, this factor is around 100.

\item
The $J/\psi$ meson  decays with a probability of 12\% into a leptonic pair, $\mu^+ \mu^-$ or $e^+ e^-$. These are  very clean final states which emerge from quark annihilation.  It can be well measured in the detector because the $J/\psi$ is a very narrow resonance. The $\rho$ meson decays mostly into a pair of pions which can be also well measured. However, this is a strong decay, the $\rho$ width is large and the decay mechanism is more complicated than in the $J/\psi$ case. This may make this process less suitable as a probe of nuclear properties. 

\end{itemize}

\subsection{Proton radius}
To show the potential of  $J/\psi$ photoproduction as a probe of the  properties of matter we discuss here a classic nuclear physics subject,  the determination of the  proton radius.
  
The smallest sizes of the interaction region, $B_D$, are measured at HERA
 in the exclusive $J/\psi$ photoproduction process, in agreement with the dipole model expectations. 
The $t$ distribution of this process is also one of the  most precisely measured at HERA.
In the following we will determine the proton radius  using  data of this process.

The observed values of $B_D$ in the H1 and ZEUS experiments show an increase with growing energy, $W$. This increase is  parametrized by the HERA experiments as
\begin{equation} \label{eq:diffusion}
B_D(W)=b_0+4\alpha'\cdot \log(W/90 {\rm GeV})
\end{equation} 
For ZEUS 
$$
b_0=4.15\pm 0.05^{+0.30}_{-0.18} \; \; \; \; \; \;
\alpha'=0.115\pm0.018^{+0.008}_{-0.015}
$$
and for H1 
$$
b_0=4.63\pm 0.06^{+0.043}_{-0.163} \; \; \; \; \; \;
\alpha'=0.164\pm0.028^{+0.030}_{-0.030}
$$
The $W$ region extends from 30 to 170 GeV and the values of $B_D$ increase with $W$ by about 25\%,
 due  to the QCD evolution effects. 
The proton radius should be related to the value of $B_D$ without this effect, i.e. to the value of $B_D$ at  $W$  near the threshold.
%  of the $\gamma p \rightarrow J/\psi p$ reaction. 
However, it is not known whether the formulae~(\ref{eq:diffusion}) can be extrapolated below $W=30$ GeV.  
%This corresponds to  the region of higher $x$ values,   $x > 10^{-2}$, where  the reggeon exchange process could also contribute and change the $B_D(W)$ dependence. 
Therefore we  evaluate the proton radius from the lowest observed value of $B_D$ at $W=30$ GeV. This value of $W$ corresponds also to the $x$ value, $x=10^{-2}$, up to which the dipole picture was successfully tested.

At  $W=30$ GeV the size of the interaction region is $B_D= 3.64$ GeV$^{-2}$ for ZEUS and $B_D= 3.91$ GeV$^{-2}$ for H1.    
We can combine these values to 
$$
B_D(W=30\; {\rm GeV})= 3.78 \pm 0.3 \; {\rm GeV}^{-2}
$$
As error we take the difference between the ZEUS and H1 values.

The proton size is related to the value of $B_G$ which is smaller than $B_D$, due to the contribution of the size of the $J/\psi$. The difference at $W=30$ GeV is  
$B_D -B_G=\Delta B=0.6$ GeV$^{-2}$~\cite{KMW} and is almost independent of $W$. The theoretical error can be estimated from evaluating $B_G$ with two different wave functions
as $\Delta B= \pm 0.2$ GeV$^{-2}$~\cite{KMW}.

The corresponding proton gaussian width is
$B_G =  3.18 \pm 0.4 \; {\rm GeV}^{-2}$, where we added the theoretical and experimental errors in quadrature. 
The transverse proton radius is then
$$
\sqrt{<b^2>}  =\sqrt{\int d^2 \vec{b} \, b^2 T_G(b)}  = \sqrt{2\cdot B_G} = 0.50 \pm 0.03\; {\rm fm} 
$$   
The proton radius is usually determined from the electromagnetic charge form factor,
$$
G_E(t) = 1+ \frac{1}{6}<r^2_p>t +  O(t^2) 
$$
and
$$
<r^2_p> = 6\cdot \frac{dG_E(t)}{dt}|_{t=0}  
$$
This is the 3-dimensional proton radius, the transverse proton radius is given by
$$
<b^2_p> = 4\cdot \frac{dG_E(t)}{dt}|_{t=0}  
$$  
Therefore the transverse proton radius is related to the 3-dimensional proton radius by $<b^2_p> =2<r^2_p>/3$~\cite{StrWeis,Diehl}. Guided by this we obtain the 3-dimensional proton radius as measured by the exclusive $J/\psi$ photoproduction 
$$
\sqrt{<r^2_{2g}>} = \sqrt{3\cdot B_G} =0.61\pm 0.04 \; {\rm fm} 
$$   
We call this radius $r_{2g}$   to indicate that it is determined in the two gluon exchange process. 

%\begin{figure}
%  \centering
%  \includegraphics[width=0.45\textwidth]{dsp.eps}
%  \caption{The $t$ measurement for the $J/\psi$ photoproduction.}
%  \label{fig:dsp_jpsi}
%\end{figure}

The two gluon proton radius, $r_{2g}$, is much smaller than the charged proton radius determined from the electromagnetic form factor $G_E$, $r_p=0.875\pm 0.007$ fm.  One could argue that it is more appropriate to  compare the two gluon proton radius to the proton radius which  is determined from the Dirac form factor $F_1$, $r_F=0.81$ fm,  instead of $G_E$, as proposed in ref~\cite{Diehl}. This value is smaller than the standard charged proton radius because the Dirac form factor describes only the spin non-flip interactions (in the infinite momentum frame). Nevertheless, the spin preserving proton radius is still substantially larger than the two gluon proton radius.   

It is expected that the value of the proton radius is process dependent because  the current which tests the proton has itself a structure which depends on its quantum numbers~\cite{Meissner}.
The smallest proton radius is determined from the axial form factor $G_A$ measured in  neutrino scattering $\nu p\rightarrow \mu p$. It is called the axial radius and has the value of $r_A = 0.675\pm 0.02$ fm~\cite{Meissner}. The smallness of this radius can be attributed to the fact that the axial current is not coupling to the pion cloud surrounding the  bulk of the proton~\cite{StrWeis}. It is interesting to observe that the two gluon radius is still smaller  than the axial radius in spite of the fact that the two gluon intermediate state couples as well to the bulk of the proton  as to the surrounding pions.

\subsection{Nuclear structure}

The measurements of diffractive processes on nuclei could become an important source of information on nuclear structure and high density QCD. 
The interaction of  a dipole with a  nucleus 
can be viewed as  a sum of dipole scatterings of the  nucleons forming the nucleus. 
The size of  the charmed dipole in  elastic  $J/\psi$  scattering  is  around 0.15  fm~\cite{KT}, i.e., it is much smaller than the nucleon radius. It is therefore  possible that dipoles interact with smaller objects than nucleons; e.g., with constituent quarks or hot spots. 
Nevertheless, for the sake of example, we take here the conventional point of view and assume that the nucleus is build out of nucleons and that dipoles scatter  on the ensemble of  nucleons.   
We  discuss the dipole-nucleus  scattering   with two  examples to show the potential of   possible nuclear investigations.  
To simplify the discussion multiple scattering effects will be ignored justified by the  small size of the $J/\psi$ dipole, see also Section 3.3.

In the first example we consider  elastic scattering on the nucleus~\cite{KT,KLV}.
This is a coherent scattering process since the nucleus remains in its ground state.
Ignoring  possible multiple scattering  effects,
the dipole scattering amplitude for a given configuration of nucleons $\{\vec{b}_i\}$, using eq.~\ref{eq:dxs} and eq.~\ref{eq:Tofb}
%\footnote{Note that the dipole cross section is also an amplitude,
 % eq.~\ref{eq:totsiggp} and~\ref{eq:difsiggp}}
is  given by             
\begin{equation}
    \frac{d \sigma^A_{q\bar{q}}} {d^2b}  = \sigma_p \sum_{i=1}^{A}
          \frac{e^{-(\vec{b}-\vec{b_i})^2/2B_p}}{2\pi B_p} ,
\label{eq:toy0}
\end{equation}
where $A$ denotes the number of nucleons in the nucleus, $B_p$ is the diffractive slope of the proton and  $\sigma_p$ is the total proton dipole cross section. 
 The Fourier transform of this  amplitude  is   
\begin{equation}
\int d^2be^{-i\vec{b}\cdot \vec{\Delta}} \,   \frac{d \sigma^A_{q\bar{q}}} {d^2b}  
= \sigma_p \sum_{i=1}^{A} e^{-i\vec{b_i}\cdot \vec{\Delta}} \cdot
          e^{-B_p \cdot \Delta^2/2} .
\label{eq:toyinc}
\end{equation}
Here we changed the integration variable from $\vec{b}$ to $\vec{b}-\vec{b_i}$
while integrating each term of the sum.  We obtain the matrix element for the elastic scattering $(q\bar q)+A_0\rightarrow (q\bar q)+A_0$ by averaging over all configurations of the nucleus ground state\footnote{To simplify the notation we write $\Psi(...)$ as a function of the transverse variables, $\vec{b}_i$, only; i.e. we assume that the  longitudinal dimensions, $z_i$, are already integrated out}. 
\begin{equation}
-iA^{q\bar q}_{A_0\rightarrow A_0}= \sigma_p   e^{-B_p \cdot \Delta^2/2} \,\sum_{i=1}^{A} 
	\int d^2 \vec{b}_1...d^2 \vec{b}_A \Psi^*_{A_0}(\vec{b}_1...\vec{b}_A)\Psi_{A_0}(\vec{b}_1...\vec{b}_A)
\cdot e^{-i\vec{b_i}\cdot \vec{\Delta}}  
\label{eq:am-el}
\end{equation}
We define the single nucleon distribution as
\begin{equation}
\int d^2 \vec{b}_2...d^2 \vec{b}_{A} d^2  \Psi^*_{A_0}(\vec{b}_1...\vec{b}_A)\Psi_{A_0}(\vec{b}_1...\vec{b}_A)= T_{A}(b_1)
\label{eq:am-el}
\end{equation}
with normalization $\int d^2b_1 T_{A}(b_1)=1$.
Because the wave function for protons and neutrons is completely antisymmetric and the  difference between proton and neutron is presumably small we assume that  $T_{A}(b_1)=T_{A}(b_i)$.  The cross section for the coherent dipole scattering is then given by 
\begin{equation}
  \frac{d\sigma^{q\bar q}_{A_0 \rightarrow A_0}} {dt} = \frac{A^2 \sigma_p^2}{16\pi} e^{-B_p\cdot \Delta^2}  \cdot 
  \left| \int d^2b \, T_{A}(b)e^{-i\vec{b}\cdot \vec{\Delta}} \right|^2,
\label{eq:coh}
\end{equation}
i.e. it is proportional to the square of the Fourier transform of the single nucleon distribution.

\begin{figure}
  \centering
  \includegraphics[width=0.45\textwidth]{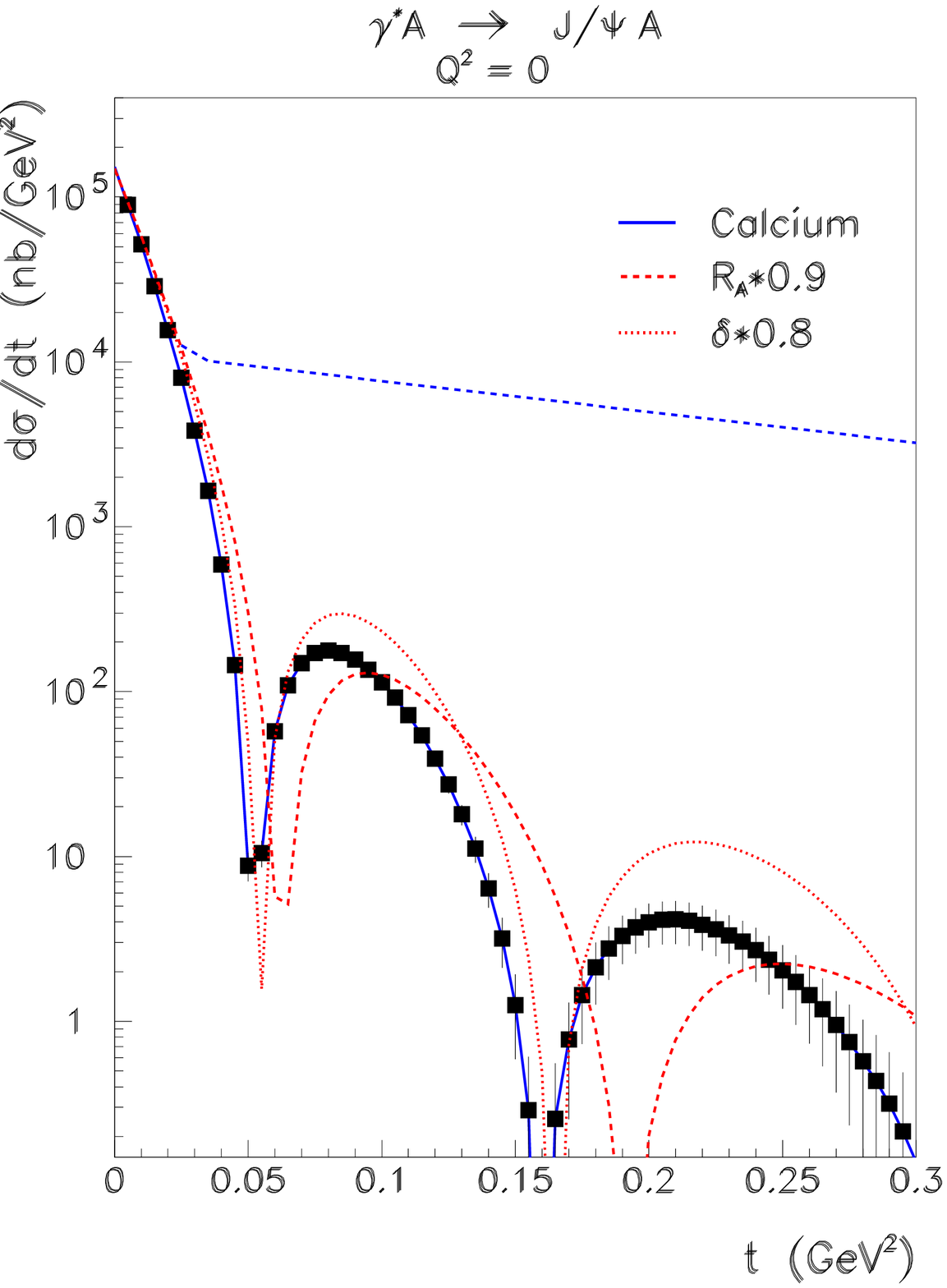}
  \includegraphics[width=0.45\textwidth]{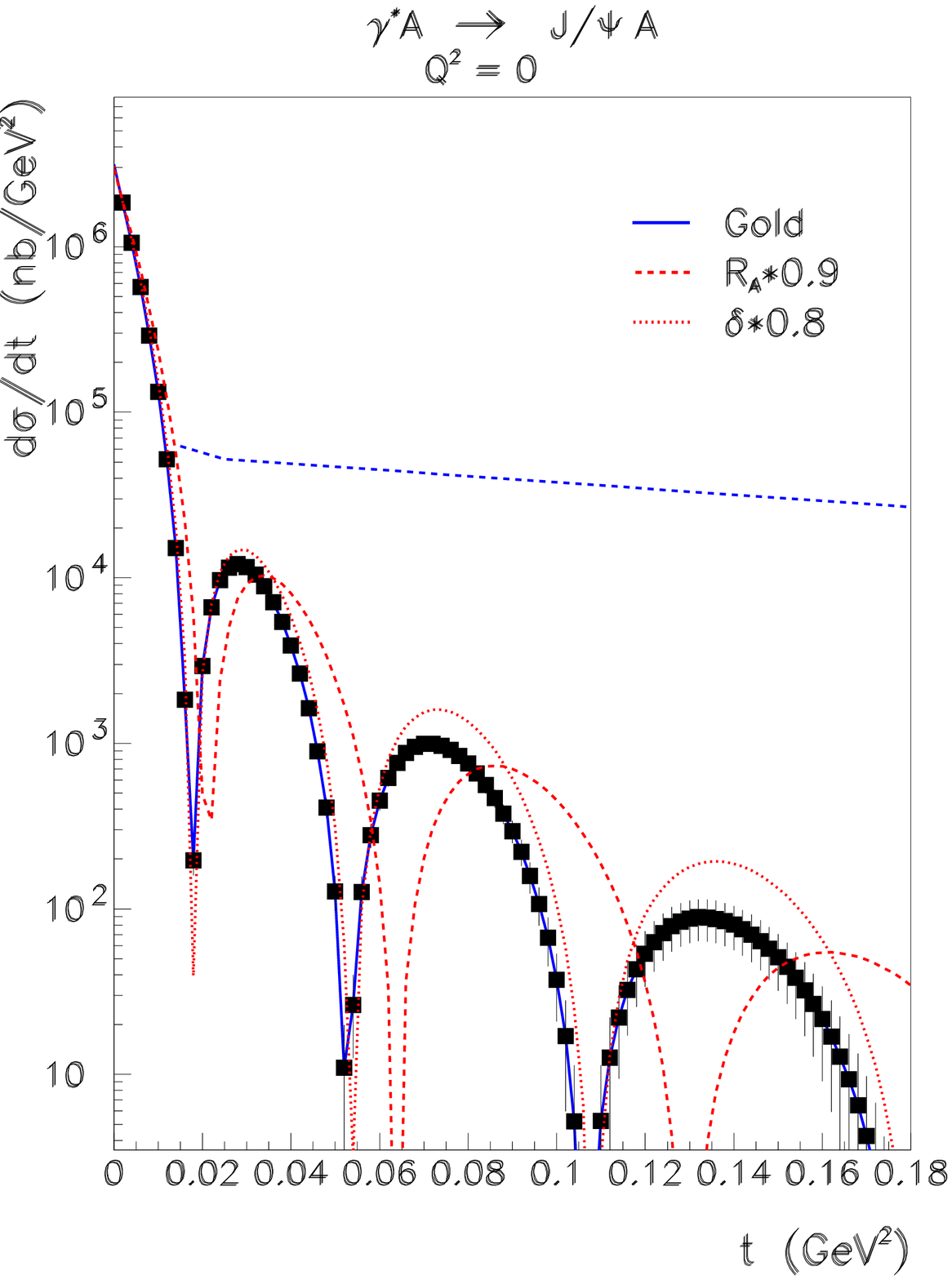}
  \caption{The prediction of the dipole model for the  $t$ distribution of  coherent  $J/\psi$ photoproduction on nuclei assuming that the single nucleon distribution can be identified with the Woods-Saxon distribution given in Appendix. The statistical errors of the simulated measurements are based on the assumed collected sample of $10^6$ events. The upper dashed line shows the sum of the coherent and incoherent process in case of no correlations.}
  \label{fig:nuclear}
\end{figure}

%  $\sigma^{\gamma A_0\rightarrow J/\psi A_0}$
 The  cross sections for  coherent $J/\psi$ scattering on nuclei, computed in the dipole model by averaging the dipole cross section, eq.~\ref{eq:coh}, over the incoming photon and outgoing $J/\psi$ wave functions,
  are shown  in Fig.~\ref{fig:nuclear} as a function of $t$. Note that  in the nuclear case, the size of the  vector meson can be ignored. We assumed here that the single nucleon distribution can be identified with the Woods-Saxon distribution, see Appendix.
The diffractive slope at $t=0$ depends on the size of the system. The figure shows for small $t\sim 1/R_A^2$ a very steep $t$ dependence, 
$\sim \exp(-tR^2_A/3)$, and then several diffractive minima.
For larger nuclei the Woods-Saxon shapes
are approximately similar to a box with a size $R_A$.  The Fourier transform of a box is given by the Bessel function $J_1(R_A\cdot \Delta)$ 
which has zeros at $R_A\cdot \Delta=3.8,\, 7.0,\, 10.2....$\footnote{This is like the Frauenhofer diffraction on a circular aperture.}. This gives the approximate positions of the  diffractive minima 
 for the exclusive  $J/\psi$ photoproduction on Calcium (A=40) and Gold (A=197) seen in Fig.~\ref{fig:nuclear}.  
 The parameters of the Woods-Saxon distribution,   nuclear radius and  skin depth, were determined mainly by scattering of  the charged matter and can be pretty different when measured in dipole interactions.
% It is well possible that, as in the proton case, the nuclear radius  can  be quite different when measured by  elastic $J/\psi$ scattering. The  skin depth has to be very different because the density of gluonic matter is different than that of the charged matter. 
We therefore plotted in Fig.~\ref{fig:nuclear} the predictions computed with slightly altered values of these parameters. 

In the second example we evaluate  incoherent  dipole scattering~\cite{privMuel},~\cite{KLV,KLMV, KLMVe}. We start with the  quasi-elastic scattering of a dipole on a nucleus
$$(q\bar q)\, A_0\rightarrow  \sum_n (q\bar q)\, A_n,$$
where $A_n$ can be either a ground state or any excited nuclear state or any breakup of the nucleus into nucleons or nucleonic fragments. Pion and other hadronic production is not  allowed. Like in the elastic case, the  matrix element for  transition to a state $A_n$ is 
\begin{equation}
-iA^{q\bar q}_{A_0\rightarrow A_n}= \sigma_p   e^{-B_p \cdot \Delta^2/2} \,\sum_{i=1}^{A} 
\int d^2 \vec{b}_1...d^2 \vec{b}_A   \Psi^*_{A_n}(\vec{b_1}...\vec{b_A})\Psi_{A_0}(\vec{b_1}...\vec{b_A})
\cdot e^{-i\vec{b_i}\cdot \vec{\Delta}}  
\label{eq:am-incl}
\end{equation}
The quasi-elastic dipole cross section
\begin{eqnarray*}
  \sum_n \frac{d\sigma^{q\bar q}_{A_0 \rightarrow A_n}} {dt} & = & \frac{1}{16\pi} \sum_n \left|A^{q\bar q}_{A_0\rightarrow A_n}
  \right|^2\\
 & =&  \frac{\sigma_p^2}{16\pi} e^{-B_p\Delta^2} 
  \sum_i^A \sum_j^A \int d^2 \vec{b}_1...d^2 \vec{b}_A  d^2 \vec{b'}_1...d^2 \vec{b'}_A 
    \Psi^*_{A_0}(\vec{b'_1}...\vec{b'_A}) \cdot \\
  &  &  \sum_n \Psi_{A_n}(\vec{b'_1}...\vec{b'_A})\Psi^*_{A_n}(\vec{b_1}...\vec{b_A})
\,    \Psi_{A_0}(\vec{b_1}...\vec{b_A})
\cdot e^{-i(\vec{b}_i-\vec{b'_j})\cdot \vec{\Delta}} 
\label{eq:inc2}
\end{eqnarray*}
can be evaluated using the completeness relation
\begin{equation}
\sum_n \Psi_{A_n}(\vec{b'_1}...\vec{b'_A})\Psi^*_{A_n}(\vec{b_1}...\vec{b_A})= \delta(\vec{b}_1-\vec{b'}_1)
... \delta(\vec{b}_A-\vec{b'}_A)
\label{eq:am-inc3}
\end{equation}
giving
\begin{equation}
  \sum_n \frac{d\sigma^{q\bar q}_{A_0 \rightarrow A_n}} {dt} 
  =  \frac{\sigma_p^2}{16\pi} e^{-B_p\Delta^2} 
  \sum_i^A \sum_j^A \int d^2 \vec{b}_1...d^2 \vec{b}_A  
    \Psi^*_{A_0}(\vec{b_1}...\vec{b_A}) \cdot 
    \Psi_{A_0}(\vec{b_1}...\vec{b_A})
\cdot e^{-i(\vec{b}_i-\vec{b_j})\cdot \vec{\Delta}} 
\label{eq:inc2}
\end{equation}
Defining the two nucleon distribution as
\begin{equation}
\int d^2 \vec{b}_3...d^2 \vec{b}_{A} d^2  \Psi^*_{A_0}(\vec{b}_1...\vec{b}_A)\Psi_{A_0}(\vec{b}_1...\vec{b}_A)= T^{(2)}_{A}(\vec{b}_1,\vec{b}_2).
\label{eq:am-2n}
\end{equation}
with normalization $\int d^2b_1 d^2b_2 T_A^{(2)}(\vec{b}_1,\vec{b}_2)=1$, we obtain 
\begin{equation}
  \sum_n \frac{d\sigma^{q\bar q}_{A_0 \rightarrow A_n}} {dt} 
  =  \frac{\sigma_p^2}{16\pi} e^{-B_p\Delta^2} 
  \left[ A + A(A-1)\int d^2 \vec{b}_1d^2 \vec{b}_2  T^{(2)}_A(\vec{b}_1,\vec{b}_2) 
\cdot e^{-i(\vec{b}_1-\vec{b_2})\cdot \vec{\Delta}}  \right].
\label{eq:inc3}
\end{equation}
The first term in the square brackets, proportional to $A$, emerges from the summation of terms with  $i=j$ in  eq.~\ref{eq:inc2}. The second term, proportional to $A(A-1)$, is obtained using  the antisymmetry property of the wave function and the assumption that the difference between protons and neutrons can be neglected, $T^{(2)}_A(b_1,b_2)=T^{(2)}_A(b_i,b_j)$ for $i\neq j$.

The incoherent dipole cross section
%\footnote{The sum of the coherent, eq.~\ref{eq:coh}, and incoherent cross sections, eq.~\ref{eq:incoh}, in the limit  where nucleon correlations are neglected gives the cross section which was labeled as   "breakup"  in ref.~\cite{KLMV,KLV} (instead of "with beakup" as originally intended). This potentially confusing word choice was clarified in~\cite{KLMVe}.}  
is obtained by  subtracting the ground state contribution from the sum of eq.~\ref{eq:inc3}  
 \begin{eqnarray}
  \sum_{n \neq 0}\frac{d\sigma^{q\bar q}_{A_0 \rightarrow A_n}} {dt}   =   
  \frac{\sigma_p^2}{16\pi} e^{-B_p\Delta^2} 
	  \int d^2 \vec{b}_1d^2 \vec{b}_2& &  \!\!\!\!\!\!\!\!  \left\{ A\ \left(T_{A}(b_1) T_A(b_2) -T^{(2)}_A(\vec{b}_1,\vec{b}_2) 
 e^{-i(\vec{b}_1-\vec{b_2})\cdot \vec{\Delta}} \right) \right. \nonumber\\ &&
 \!\!\!\!\!\!\!\!\!\!\!\!\!\!\!\! +  \left. \, A^2 \left(T^{(2)}_A(\vec{b}_1,\vec{b}_2)-T_{A}(b_1) T_{A}(b_2)   \right) 
  e^{-i(\vec{b}_1-\vec{b_2})\cdot \vec{\Delta}} 
 \right\}
\label{eq:incoh}
\end{eqnarray}  
When the momentum transfer $\Delta\rightarrow 0$ the incoherent dipole cross section, eq.~\ref{eq:incoh}, goes to zero reflecting the fact that  without a transfer of transverse momentum to the nucleus excited states cannot be produced. At larger values of the momentum transfer,  $| \vec{\Delta} | > 200  $ MeV, the contribution of the second term in the upper bracket of eq.~\ref{eq:incoh}, $ \int d^2 \vec{b}_1d^2 \vec{b}_2  \, A\ \left(T^{(2)}_A(\vec{b}_1,\vec{b}_2) 
 e^{-i(\vec{b}_1-\vec{b_2})\cdot \vec{\Delta}} \right) $, starts to be relatively smaller than that of the first term because  the oscillatory factor  suppresses the contributions with $|\vec{b}_1-\vec{b}_2| > 1$~fm. The contribution of the second term grows then as $A^{1/3}$ and can be neglected for larger nuclei compared to the contribution of the first term which grows as $A$.
 Thus when $| \vec{\Delta} | > 200  $ MeV 
\begin{eqnarray}
  \sum_{n \neq 0}\frac{d\sigma^{q\bar q}_{A_0 \rightarrow A_n}} {dt} &   = &  
  \frac{Ad\sigma^{q\bar q}_{p\rightarrow p}}{dt} \nonumber \\ &+&
  \frac{\sigma_p^2}{16\pi} e^{-B_p\Delta^2} 
	A^2   \!\!  \int d^2 \vec{b}_1d^2 \vec{b}_2    
    \left(T^{(2)}_A(\vec{b}_1,\vec{b}_2) -T_{A}(b_1) T_{A}(b_2)   \right) 
  e^{-i(\vec{b}_1-\vec{b_2})\cdot \vec{\Delta}} 
\label{eq:incoh1}
\end{eqnarray}  
and the deviation from the elastic nucleon cross section result is due to two-body correlations only. 

Experimentally we expect to be able to distinguish cases where the nucleus remains intact and cases where the nucleus breaks up.  In the nuclear breakup process, there are around $0.3\sqrt{A}$ free neutrons and $0.2\sqrt{A}$ protons in the final state ~\cite{Ranft}, as well as other fragments.  These particles and fragments have high momenta and different charge to mass ratios than the nuclear beam, and should therefore be measurable in specialized detectors.  However, we do not have a one-to-one correspondence between an intact nucleus and a coherent scattering process since incoherent processes can lead to an intact nucleus in an excited state.  It remains to be determined how well excited states of the nucleus can be identified, measured and   possibly statistically subtracted to extract the fully separated coherent and incoherent processes.

%The coherent and incoherent processes are presumably  experimentally well distinguishable because in the coherent case no additional particles are emitted whereas in the incoherent case   free protons, neutrons or photons are emitted.
%In the  nucleus breakup process there are around  $0.3\cdot \sqrt{A}$ free neutrons  and  $0.2\cdot \sqrt{A}$ protons in the final state ~\cite{Ranft}. Their momenta are close to the ion momentum and  can be 
%  measured in the specially build forward detectors with high efficiency.   When nucleus gets excited without breakup  low energy photons are  emitted which, because of the Lorentz boost, are detectable in the forward photon counters. 

%The high  precision with which the momenta of the final particles can be measured could allow to reconstruct the invariant mass of the nucleonic system with a precision of a few MeV. This could be sufficient to distinguish the excited nucleonic system from the nucleon in its ground state. In addition,   

% Since the $t$-distributions sense the transverse extension of the forces a measurement of the $t$ distributions  correlated with the number and the momenta of the break up neutrons and protons can become an invaluable source of information about the nuclear forces~\cite{StriVen}.  

\subsection{Saturation} 

One of the main results of HERA is the observation that the gluon density increases quickly with decreasing $x$. This suggests that at some small $x$ the dipole could undergo multiple interactions.
The degree of saturation is characterized by the  size of the dipole $r_S$ which,
at a given $x$, starts to interact multiple times. The dipole size $r_S$ is defined, by convention~\cite{KMW}, via the relation\footnote{
From the unitarity limit the highest value of a dipole cross section is $d\sigma/d^2b=2$, see eq.~
\ref{eq:xdip}.}
\begin{eqnarray}
\frac{d\sigma_{q\bar q}(x,r_S,b)}{d^2b} = 2(1-\exp(-1/2))\approx 0.8.
\label{eq:dxsc}
\end{eqnarray}
 The saturation scale is defined then as $Q_S^2=2/r_S^2$ and is a function of $x$.
A high value of the saturation scale means that  gluon density is so high that even small dipoles
 interact many times. 

\begin{figure}[t]
\centering
\includegraphics[height=8cm,width=8cm]{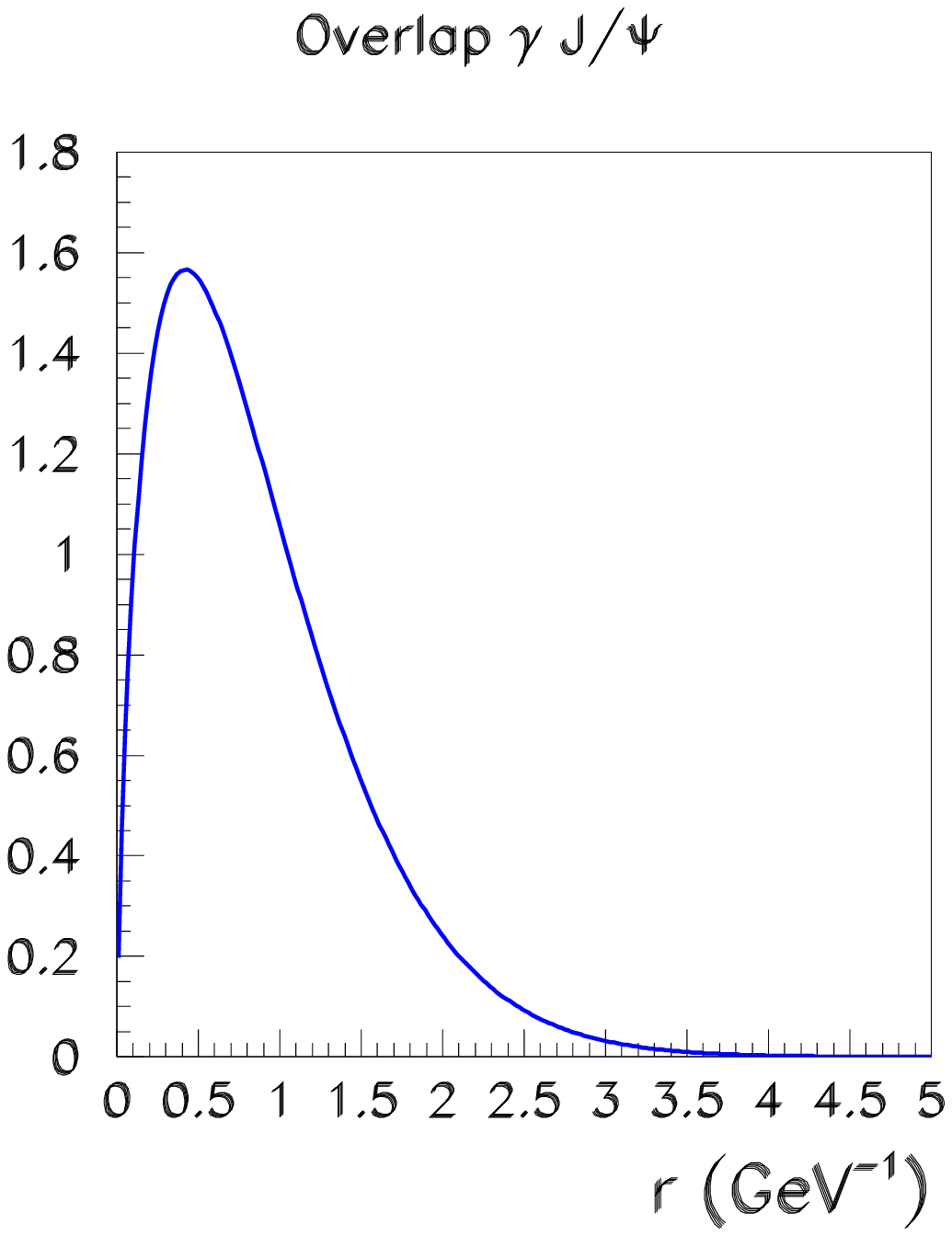}
\includegraphics[height=8cm,width=8cm]{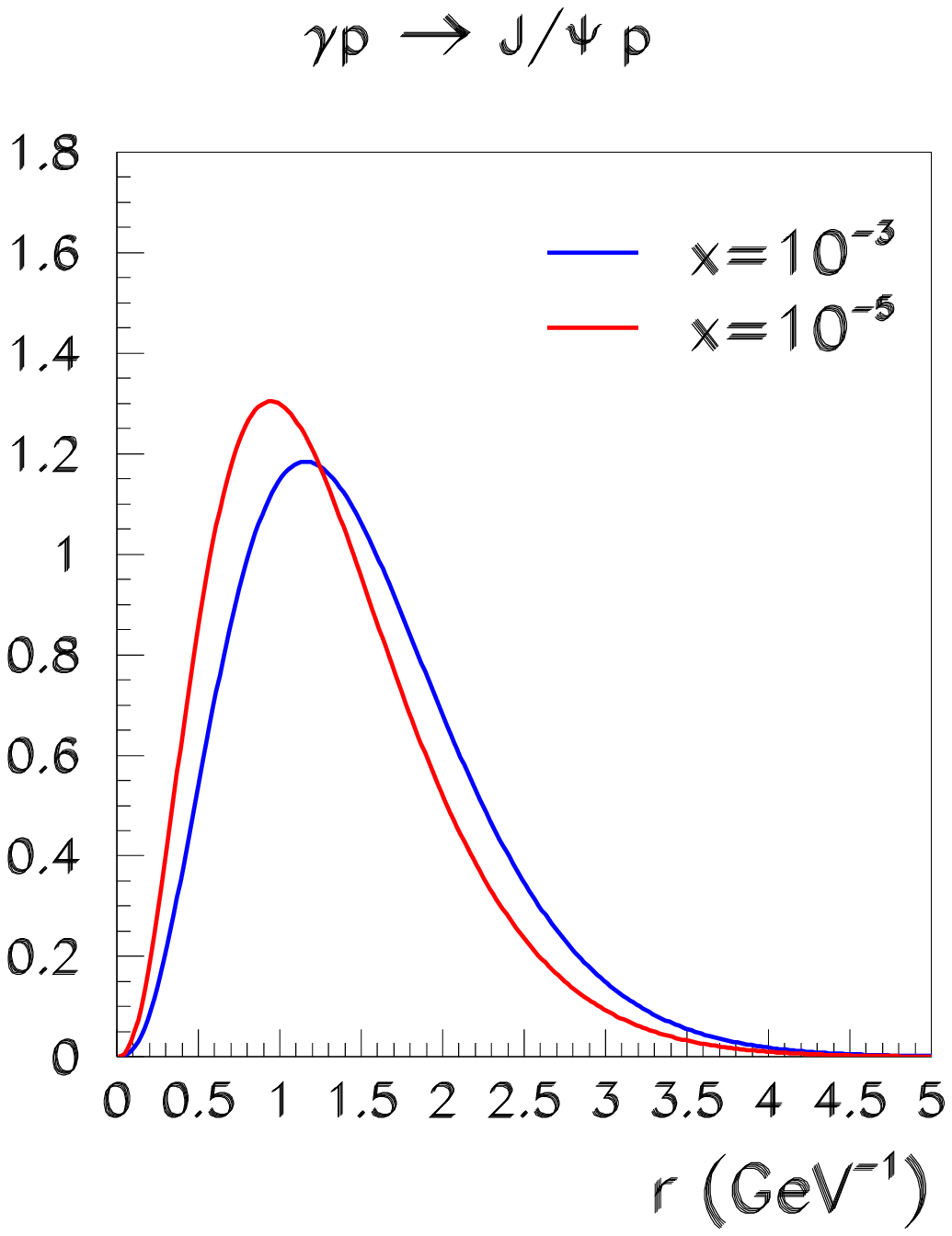}
\caption{\label{fig:ampel}
 The distribution of the $J/\psi$ dipole sizes; {\it Left:} as given by the $\gamma J/\psi$ overlap function; {\it Right:} as given by the   photoproduction amplitude.}
\end{figure}

\begin{figure}[t]
\centering
\includegraphics[height=9cm,width=9cm]{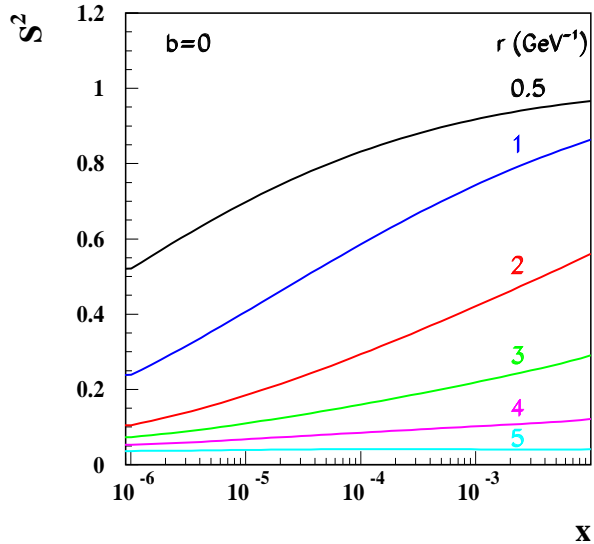}
\caption{\label{fig:s2mat}
 The survival probability $S^2$ as a function of $x$ for various dipole sizes.}
\end{figure}

In various analyses of  HERA data the saturation scale,  in the proton center,   was found to be $Q_S^2 \approx 0.5$ GeV$^2$ at $x\approx 10^{-3}$, i.e. in the EIC range. In the LHeC range, which extends  to $x \approx 10^{-5}$, the saturation scale could reach $Q_S^2\approx 2$ GeV$^2$~\cite{KMW}.  
The saturation scale determined in  the inclusive $\gamma^* p$ reaction   should be compared to the  scale of $J/\psi$  photoproduction   given by the effective size of  the meson in this reaction. This size is determined by the overlap of the photon and $J/\psi$  wave functions and by the dynamical effects.     Fig.~\ref{fig:ampel}{\it Left} shows the distribution of the dipole sizes  selected by the overlap of the photon and $J/\psi$ wave functions defined as 
$r\int dz \Psi^*_{J/\psi} \Psi$. The median value of this distribution is $r_{med}^o = 0.7$ GeV$^{-1}$, i.e. around 0.15 fm.
The contribution of very small dipoles, in the $\gamma p$ reaction,  is suppressed by the dynamic of the reaction,  because  the dipole  cross section is proportional to $r^2$ at small $r$, eq.~\ref{eq:dxs}. The distribution of the dipole sizes selected by the amplitude, eq.\ref{eq:amvecm},  is shown in
Fig.~\ref{fig:ampel}{\it Right}. The median value of  this distribution, $r^a_{med}$, was found \footnote{The evaluation was performed  in the b-Sat model with the' Boosted Gaussian' wave function , at $t=0$} to be between  1.2 and 1.4 GeV$^{-1}$, depending somewhat on $x$.
In the $\gamma p$ reaction the effective scale characteristic for $J/\psi$ photoproduction is then $Q_{eff}^2 =2/(r^a_{med})^2\approx 1-1.5$ GeV$^2$, depending on $x$. 
The saturation scale at EIC is sizably smaller then $Q_{eff}^2$,  so that saturation effects  are not expected to be large. However, the measurement on nuclei could enhance the saturation scale substantially, as discussed in ref.~\cite{KLV}. Therefore, the
  high  precision which can be achieved in the measurement of $J/\psi$ exclusive photoproduction makes this process  interesting as an alternative testbed of  saturation effects at EIC. 
   At LHeC  saturation effects should be clearly visible in the scattering on proton and presumably very strong in nuclear reactions.

In ref.~\cite{MSM} it was proposed to investigate  the effects of saturation in a systematic way
 by extracting from  data   $S^2(b)$, the square of the $S$ matrix. The $S$ matrix is directly connected to the dipole cross section by $d\sigma_{qq}/d^2b=2[1-\Re S(b)]$ as shown in  eq.~\ref{eq:difsiggp}.
The $S$ matrix can  also be used to define the saturation condition because its square has a meaning of the survival probability;  i.e., the probability for no interaction. The equivalent definition to eq.~\ref{eq:dxsc} then reads
 $S^2=e^{-1}\approx 0.37$. 
 Fig.~\ref{fig:s2mat} shows the survival probability for dipoles of different sizes scattering on the proton, at $b=0$,
 where the gluon density reaches its maximum~\cite{KT}.
 
To reconstruct the $S$ matrix at $b=0$  for protons it is necessary to measure the $t$ distribution up to about $|t|\approx 2$ GeV$^2$~\footnote{In case of nuclei the measurement will require  a much smaller $t$ range than in the proton case.}.  At HERA this measurement could not  be performed because the measurement of the $t$ distribution had large systematic errors for $|t|>1$ GeV$^2$ and a low statistical significance. The systematic uncertainties  which plagued HERA experiments can be avoided in the specially designed experiment discussed below. In addition it should be possible to increase the collected number of events by around two orders of magnitude in comparison to HERA which should allow to reach the $|t|=2$ GeV$^2$ region in the proton measurement.

\section{Production Cross Section}
In this section, we review the production cross section for the photoproduction of $J/\psi$ and discuss the kinematic range where high precision measurements should be possible. The $ep \rightarrow ep J/\psi$ cross section can be written as
$$\frac{d^2\sigma}{dW^2dQ^2}=\frac{\alpha}{2\pi}\frac{1}{sQ^2}
\left[ \left( \frac{1+(1-y)^2}{y} - \frac{2(1-y)}{y}\frac{Q^2_{min}}{Q^2}\right)
\cdot \sigma_T^{\gamma p}(W^2,Q^2) + \frac{2(1-y)}{y}\cdot \sigma_L^{\gamma p}(W^2,Q^2) \right]$$
where 
$$Q^2_{min}=\frac{m_e^2y^2}{1-y} $$
and $y$ is the inelasticity.
We are interested in very small $Q^2$, e.g., $Q^2<10^{-2}$~GeV$^2$, and we can assume that in this range , the photoproduction cross section is independent of $Q^2$.  
We also assume that the longitudinal cross section is negligible since we are dealing with almost real photons. Then integration gives
$$\frac{d\sigma}{dW^2}=\frac{\alpha}{2\pi}\frac{1}{s}
\left[ \frac{1+(1-y)^2}{y}\ln\frac{Q^2_{max}}{Q^2_{min}} - \frac{2(1-y)}{y}\left(1-\frac{Q^2_{min}}{Q^2_{max}} \right) \right] \cdot \sigma^{\gamma p}(W^2) \; .$$

From the ZEUS data~\cite{ZEUSjpsi}, we have
$$\sigma^{\gamma p \rightarrow J/\psi p}(W^2) \approx 75 {\rm nb} \cdot \left( \frac{W^2}{8100} \right)^{0.35} \; .$$
Writing this in terms of $y$, we have
$$\sigma^{\gamma p \rightarrow J/\psi p} = 75~{\rm nb} \left(\frac{s}{8100}\right)^{0.35}y^{0.35} $$
where $W$ is in GeV and $s$ is in GeV$^2$.  This form is valid at the large $W$ measured at HERA.  An extrapolation to lower values of $W$ is in agreement with measurements performed by E401~\cite{E401}, so we will assume this form for all $W$.

We consider two different values of $Q^2_{max}$.  In one case, we assume that we do not measure the scattered electron precisely, and we can only limit the $Q^2$.  For this case, we take $Q^2_{max}=10^{-4}$~GeV$^2$ so that the maximum $p_T$ from the electron is $10$~MeV.  In the second case, we assume we do measure the scattered electron well, so we can afford to go to higher $Q^2$, and we take $Q^2_{max}=10^{-2}$~GeV$^2$.  In the  $Q^2$ range considered here, the scattered electron energy is given by $E'=(1-y)E_e$.

High precision measurements of the $J/\psi$ decay products are assumed to be made in a central detector, such as a thin-walled time projection chamber, located in a strong solenoidal magnetic field.  The detector parameters are discussed in the next section.  To measure the $J/\psi$ decay products with good acceptance in the central detector,  we require that the $J/\psi$ has limited boost.  For simplicity, we take $Q^2=t=0$ to see the limits on $y$~\cite{Mellado}. Conservation of energy gives

$$E_e+E_p=(1-x)E_p + E_e'+E_V$$
where $E_e,E_p$ are the incoming beam energies, and $E_e',E_V$ are the scattered electron and vector meson energies.  This can be rewritten as
$$x=\frac{E_V-yE_e}{E_p}\; .$$
We also have the requirement:

$$M_V^2=(xp+q)^2\approx2xp\cdot q\approx sxy$$
or
$$x=\frac{M_V^2}{sy}\;\; .$$
Putting these expressions together gives
$$E_V\approx y\cdot E_e+\frac{M_V^2}{4yE_e}$$
where $M_V$ is the mass of the $J/\psi$.  This leads to the following constraints:
\begin{eqnarray*}
y_{max}&=&min\left[1,\frac{E_V+P_V}{2E_e}\right] \\
y_{min}&=&max\left[0,\frac{E_V-P_V}{2E_e}\right] 
\end{eqnarray*}
or in terms of $W^2$,
\begin{eqnarray*}
W^2_{max}&=&sy_{max} \\
W^2_{min}&=&sy_{min} \; .
\end{eqnarray*}

We now integrate the differential cross section given above in this $W^2$ range, with the result:
$$\sigma({\rm visible}) =A\frac{\alpha}{2\pi}\left(\frac{s}{8100}\right)^b \left[
\frac{2(c-1)}{b}y^b+\frac{2(1-c)}{1+b}y^{1+b}+\frac{c}{2+b}y^{2+b}\right]_{y_{min}}^{y_{max}}$$
where we have 
\begin{eqnarray*}
A & = & 75. \, {\rm nb} \\
b & = & 0.35 \\
c & = & \ln \frac{Q^2_{max}}{Q^2_{min}}  \; .\\
\end{eqnarray*}

\begin{figure}[htbp] %  figure placement: here, top, bottom, or page
   \centering
   \includegraphics[width=2.8in]{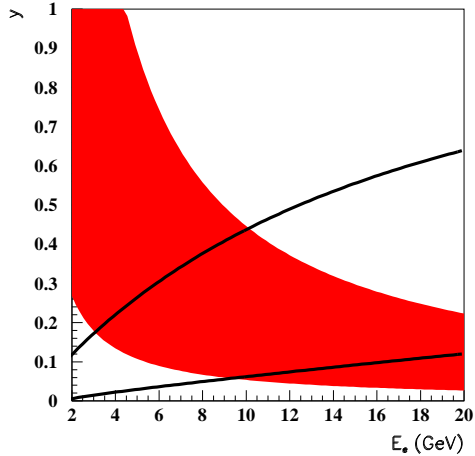} 
   \caption{The $y$ range with good acceptance for reconstructing $J/\psi$ decays in the central region is plotted as a function of the electron beam energy (shaded area) at $E_p=100$ GeV.  The two curves indicate the minimum $y$ values for detecting the beam electron scattered at $0$ degrees assuming (upper curve) that a $1$~Tm dipole is placed near the IP and that the scattered electron should be at least $5$~cm from the beam $5$~m downstream of the dipole, and (lower curve) that a $5$~Tm dipole is placed near the IP and that the scattered electron should be at least $5$~cm from the beam $10$~m downstream of the dipole.}
   \label{fig:yrange}
\end{figure}

\begin{figure}[htbp] %  figure placement: here, top, bottom, or page
   \centering
   \includegraphics[width=2.8in]{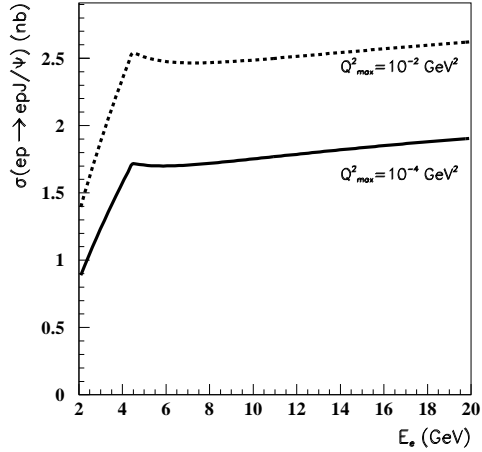} 
   \caption{Visible $J/\psi$ cross section as a function of the electron beam energy at $E_p=100$ GeV.}
   \label{fig:epcross}
\end{figure}

To see the effect of the centrality requirement, we require that the momentum of the $J/\psi$ is less than $4$~GeV, so that the decay particles are not too boosted.  We performed a scan over electron beam energy, keeping the proton beam energy fixed at $E_p=100$~GeV.  The accepted $y$ ranges are shown in Figure~\ref{fig:yrange}. The results for the visible cross section (not including branching ratio) are given in  Fig.~\ref{fig:epcross}.  Adding the branching ratio of about $6$~\% implies that $10$~fb$^{-1}$ would give $10^6$ well reconstructed $J/\psi \rightarrow \mu^+\mu^-$ events in the central part of the detector.

\section{Experimental Considerations}

In the following, we consider the requirements on the beam momentum spread and the detector requirements for the scattered electron, the measurement of the $J/\psi$, and the measurement of the scattered nucleus or proton. As discussed in Sections 2.2 and 2.3  the proton or nuclear shapes  are measured by the  deflection of  the $J/\psi$ from its original direction. The $t$ measurement will rely on an accurate determination of the transverse momentum given to the $J/\psi$  since 
$$ t = (p_i-p_f)^2\approx  -\left( (\Delta p_x)^2 + (\Delta p_y)^2 \right) $$
where $p_i, p_f$ are the  momenta of the incoming photon and outgoing $J/\psi$, and $\Delta p_x,\Delta p_y$ are the changes in the transverse momentum components during the scattering process.  As is clear from these expressions, we require knowledge of
\begin{enumerate}
\item  the outgoing $J/\psi$ momentum, and
\item the incoming and outgoing electron momenta
since 
$$  \vec{p}_{e,i} -\vec{p}_{e,f} - \vec{p}_{J/\psi} = \vec{p}_i-\vec{p}_f \;\; . $$
\end{enumerate}

Usually,  $t$ is viewed as a  change of the momentum of the incoming and the outgoing ions or protons, $t=-(\vec{p_{I,i}}-\vec{p_{I,f}})^2$. The measurement of the ion momenta is technically very difficult. Here, it is not necessary  because  $\vec{p_{I,i}}-\vec{p_{I,f}}= \vec{p}_{e,f} + \vec{p}_{J/\psi} - \vec{p}_{e,i}  $. We,
 therefore,  do not foresee a precision measurement of the ion $p_T$.  

%Note that we also do not need to require a small incoming ion transverse momentum - only that the incoming transverse momentum of the electron is small.  

Measuring the outgoing $J/\psi$ and electron (or guaranteeing that the outgoing electron has very small transverse momentum) then allows a measurement of $t$.  Nevertheless, the forward scattering region for the ions will require substantial instrumentation in order to 
\begin{enumerate}
\item
guarantee that we have an elastic scattering event. 
\item 
measure the correlation between the variable $t$ and the transverse momenta of debris from the  nuclear dissociation  in inelastic events.
\end{enumerate}
 
 We discuss the separate parts of the detector system according to their function.  Since we are dealing with photoproduction, the large majority of the events  are concentrated close to $Q^2= 0$~GeV$^2$. The scattered electron will therefore have a median transverse momentum of  only a few MeV.  As discussed below, it should be possible to measure this scattered electron over the bulk of the interesting kinematic range.  Even in cases where the electron is not detected, it will  typically only carry a very small transverse momentum.  The measurement of the $t$ distribution will therefore rely almost exclusively on a precise measurement of the $J/\psi$ decay products.

\subsection{Scattered electron}
The scattered electron will be at a very small angle to the direction of the electron beam, but can be pulled out of the beam if it has lost enough energy.  There will be a minimum value of $y$ where it is possible to see the scattered electron even if the electron is scattered at $0^{\circ}$.  For this purpose, we imagine that there will be a large-aperture dipole magnet placed near the beam interaction point which will bend the electrons (and can also act as a separator of the ion and electron beams).  The transverse separation between a beam electron and a forward scattered electron can be written as:
$$\Delta=D\left[\sin\theta_y \frac{\cos \theta_0}{\cos \theta_y} - \sin \theta_0\right]$$
where
$$\sin \theta = \frac{0.3 \cdot B \cdot L}{p} $$
with the dipole field strength $B$ measured in $T$, the length of the magnet $L$ in meters and the momentum of the electron $p$ in GeV/c.  $\theta_0$ is used to denote the scattering angle of the beam electrons after passage through the dipole, and $\theta_y$ is the scattering angle for momentum $p=(1-y)E_e$.  The distance $D$ is measured from the center of the dipole.  If detectors are placed within a few cm from the electron beam a few meters downstream of the dipole magnet, then scattered electrons at not too low $y$ can be measured.  The minimum value of $y$ required for acceptance of such a system depends on the strength of the dipole field, the distance of approach to the beamline, and the drift distance $D$.  Figure~\ref{fig:yrange} indicates the minimum $y$ values as a function of the electron beam energy for two representative sets of parameter values.  We conclude that measuring the scattered electron will be feasible over the bulk of the interesting $y$ range provided the interaction region allows for such a dipole magnet and detector arrangement.

\subsection{$J/\psi$ reconstruction}

We focus on the measurement of the $J/\psi$ via the decay into muons, $J/\psi \rightarrow \mu^+ \mu^-$.  We assume no particle identification will be necessary, since it will be easy to identify the $J/\psi$ via the invariant mass of the reconstructed state.  A continuous background from the Bethe-Heitler process $eA \rightarrow e A \mu^+ \mu^-$ will have to be handled in the analysis of the data.    
The expected resolution of the drift chamber can be estimated from the measurement precision term
$$
(\sigma_{p_t}/p_t)_{meas} = \frac{p_t \, \sigma_{r\phi}}{0.3L^2B} 
\sqrt{\frac{720}{N+4}}  
$$
and the multiple scttering contribution
$$
(\sigma_{p_t}/p_t)_{MS} = \frac{0.05}{LB\beta} \sqrt{1.43\frac{L}{X_0}}
[1+0.038\log(L/X_0)]  
$$
as
$$
\sigma_{p_t}/p_t = (\sigma_{p_t}/p_t)_{meas} \oplus (\sigma_{p_t}/p_t)_{MS}. 
$$
Here $B$ is the magnetic field in Tesla, $L$ is the lever arm in meters, $\sigma_{r\phi}$ is the spatial resolution in meters for a single point, $X_0$ is the radiation length in meter, $N$ is the number of points and $\beta$ is the velocity of the particle.  

In the $J/\psi$ photoproduction process at small $t$ the transverse momenta of the decay muons  range from very small values up to a maximum of about $2$~GeV/c.
In this momentum range, multiple scattering is a critical issue for momentum resolution.  We therefore envision a TPC-type detector with a thin inner wall as the central tracking detector.  Assuming the following parameters
\begin{enumerate}
\item outer radius $R=2$~m
\item solenoidal field $B=3.5$~T
\item gas density $X_0=450$~m
\item point resolution $\sigma=100$~$\mu$m
\item measurement $N=200$ points.
\end{enumerate}
yields a track momentum resolution~\cite{ref:standardformula}
$$ \sigma_{p_t}/p_t =0.005\cdot p_t \oplus 0.045/\beta~\%$$
which will give a $p_t$ resolution for the $J/\psi$ of typically $\leq 1$~MeV. 

%This is likely somewhat better than required, so that it should be possible to scale back on some of the requirement, such as the dimension of the TPC and  number of measurement points.  
The muons would not escape such a detector in the radial direction.  An electromagnetic calorimeter placed outside the TPC could be used to reject or measure radiated photons from the interaction vertex and from passage of the muons through the detector.  The detector design could  naturally be extended to include hadronic calorimetry, particularly in the ion direction, resulting in a general purpose detector capable of high precision measurements for all types of exclusive processes as well as inclusive cross section measurements.

\subsection{Ion forward direction}
The main requirement for the instrumentation in the forward direction of the ion beam is that dissociative and inelastic events be rejected with $\sim 100$~\% efficiency.  We envisage again a dipole magnet which can be used to separate neutral particles as well as nuclear fragments from the main beam.  This dipole could also serve as a beam separator to guide the electrons and ions to their individual beampipes.  Ion dissociation will produce different types of fragments, including neutrons and charged ions with different charge/mass ratios than the beam ions.  Neutrons can be recorded in a calorimeter at zero degrees to the beamline at the interaction point, and located many meters downstream of the interaction point.  Fragments with different charge/mass ratios will have different deflection angles in the dipole than the main beam, and can therefore be measured with tracking detectors placed close to the beamline.  In inelastic scattering events, particles with opposite charge to the beam ions can be produced, and these interactions can be vetoed using tracking detectors or calorimeters placed behind the dipole.  We therefore expect that for large nuclei, rejecting non-elastic events will be rather straightforward since typically several neutrons and charged particles will be produced. 

\subsection{Beam requirements}
The experiment discussed requires substantial instrumentation extending to many meters on either side of the interaction point.  It will be critical to avoid 'dead zones', where scattered particles could escape detection.  This will place severe restrictions on the accelerator design, and will naturally lead to limitations in luminosity.  The detector design foresees a central region with a strong solenoidal field, and dipole magnets on either side of the interaction point (perhaps +-2 meters away).  The design of these magnets will necessarily have to be done in conjunction with the accelerator group.  Issues such as synchroton radiation loads will need to be evaluated. In addition, it is critical that the electron beam transverse momentum be limited to a few MeV, since this quantity cannot be measured but enters in the calculations of $t$.

\section{Summary and Discussion}

We discussed here the physics potential of the $t$ measurements. At HERA, 
  the measurement of the $t$ distributions allows to determine  the two gluon proton radius. This radius is substantially smaller than the proton radius determined in the electromagnetic interactions, $r_{2g} < r_p$. %This may indicate that the nuclear structures seen by the two gluon interaction could  also be substantially different from the one tested by the electromagnetic current.   
    
To exemplify the potential of such measurements on  the future electron-ion colliders 
 we considered  two simple examples of coherent and incoherent $J/\psi$ dipole scattering on nuclei. In the coherent case the nucleus remains intact  while in the incoherent case  the nucleus goes into any excited state or it disintegrates into nucleonic fragments or nucleons without  production of additional hadrons. 
 %The photons,  protons and neutrons  emitted in the nucleus excitation or breakup process can be well measured in the forward detectors.  
 
  In the coherent case we computed the predictions of the dipole model assuming   that nucleons are distributed   within the nucleus according to the Woods-Saxon distribution, Fig.~\ref{fig:nuclear}, which is mainly determined by  scattering on the electric charges. It is well possible that, as in the proton case, the nuclear radius and skin parameters  can  be quite different when measured by  elastic $J/\psi$ scattering. The shape of the expected cross section and the differences between different predictions indicate that a measurement resolution of around 10 MeV for the $p_T$ of $J/\psi$  should be sufficient. This number emerges from the requirement that the first diffractive minimum should be properly resolved.

% The parameters of the Woods-Saxon distribution,   nuclear radius and  skin depth, were determined mainly by scattering on the electric charges. 
% It is well possible that, as in the proton case, the nuclear radius  can  be quite different when measured by  elastic $J/\psi$ scattering. The  skin depth has to be very different because the density of gluonic matter is different than that of the charged matter. 
%We therefore plotted in Fig.~\ref{fig:nuclear} the predictions computed with slightly altered values of these parameters. 

Incoherent $J/\psi$ scattering is equally  interesting since the very good $p_T$ resolution combined with the full acceptance detector discussed here should allow a  systematic study of the two body correlations.  The measurement of the long range nuclear correlations could require a measurement resolution of  O(1) MeV in  $p_T$  of  the $J/\psi$ decay products.  

The $t$ distribution of the nuclear breakup process will be measured in correlation with the number and momenta of the breakup protons and neutrons. This should allow to study the dissociation process as a function of the transverse momentum transfered to the nucleus.

Although we concentrated here on the measurement of  $J/\psi$ decays it should be clear that the detector which 
is optimal for measurements of   $J/\psi$ elastic scattering  will also be optimal for measurements of other  interactions, like light vector meson production and inclusive diffractive or inclusive total cross sections.
The quality of the inclusive measurements will also profit from the large acceptance  detector discussed here because the coverage of the almost entire rapidity range for charge and neutral particles  will reduce the systematic uncertainties in the $F_2$ and $F_2^D$ measurements.
 
The nuclear effects will  also be seen through the measurement of the 
 absolute value of  cross sections. The dipole model predicts these values precisely provided the scattering takes place on nucleons within the nucleus which have the same properties as  free protons.
 Any deviation   from the expected value carries information about  nuclear effects.
  For example, the total dipole proton cross section, $\sigma_p$, can be different for a nucleon in a nucleus than for a proton because a nucleon within a nucleus can have a different size than a free proton or neutron. This would change sizably the value of the nuclear dipole cross section and therefore also the values of 
the observed diffractive cross sections. By the same argument the measurement of $F_2$ on nuclei is also  determined by the nuclear properties and will lead to very interesting saturation effects as discussed in~\cite{KT,KLV}. 
Saturation is dependent on the size of the scattering objects, therefore its measurement and the absolute value of the cross section could  indicate on what objects the scattering takes place.

 The measurement of the transverse shape and correlations combined with the measurements of the total cross sections and the shadowing effects  
should allow to determine the inner structure of gluonic fields which keep the nuclei together.  
    
%This, combined with the information from the transverse shapes, should give a detailed insight into the nucleus.
\section{Conclusion}

We have shown that dipole interactions of small dipoles   with nuclei can become an important source of information on the nuclear structure. At high energies, the dipoles interact with nuclei by a well understood QCD process in which two gluons, with large transverse momenta, are exchanged. The difference of the gluon momenta can be precisely determined by measuring the transverse momentum of the elastically scattered vector meson.    The $J/\psi$ photo-production process is particularly well suited to perform this measurement  because it has the largest cross section and the best measurement precision.  The $p_T$ of $J/\psi$ can be measured  with high efficiency and a precision of O(1) MeV  using presently available techniques.  This allows to determine the $t$ distributions with a  precision O(1) MeV$^2$, starting at $t \approx 0$. The upper range of the available  $t$ values depends on the process and  can reach O(3) GeV$^2$ with the luminosity foreseen   at EIC or LHeC.  This  allows to investigate large and/or small gluonic structures which keeps the matter together. It is worth emphasizing that large structures are investigated by a highly virtual, well understood interaction because a small $p_T$ of the $J/\psi$ meson is a result of the difference of  two large transverse  gluon momenta.  Therefore, dipole measurements will provide high quality  data of basic nature which could lead to  solution of the long standing puzzle; how strong interactions form  matter.

% which together with the   Lattice Gauge Theory
%   We have therefore a chance to solve the long standing puzzle; how strong interactions are forming the matter.

\section{Acknowledgement}
The approach
to nuclear structure  described in Section 3.2 was  developed  together with T. Lappi and R. Venugopalan.   Writing this Section  we profited   from comments of Al Mueller in an essential way.  
We would like also to thank M. Diehl, V. Litvinienko, A. Spiridonov, L. Motyka and U. Schneekloth for useful comments and discussions. 
\section{Appendix}
\subsection{Derivation of the dipole representation}
\begin{figure}[htbp] %  figure placement: here, top, bottom, or page
   \centering
   \includegraphics[width=3.2in]{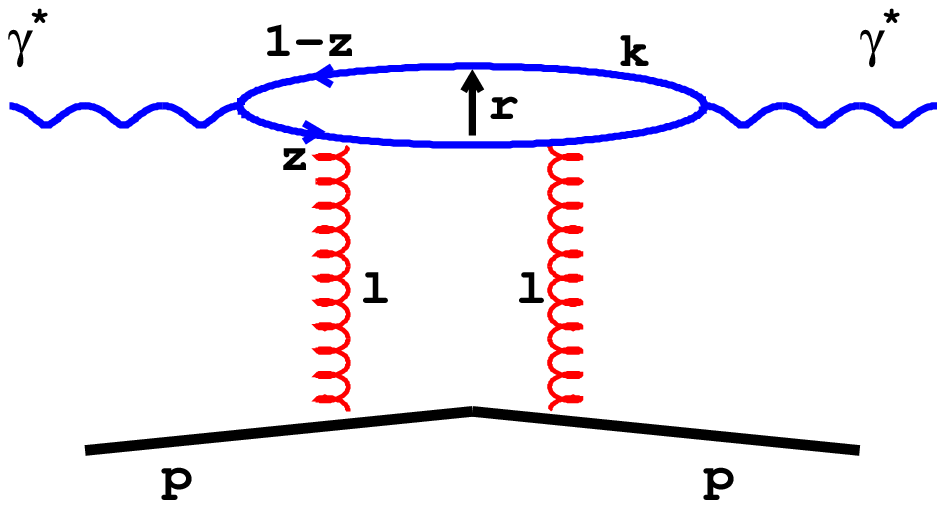} 
   \includegraphics[width=3.2in]{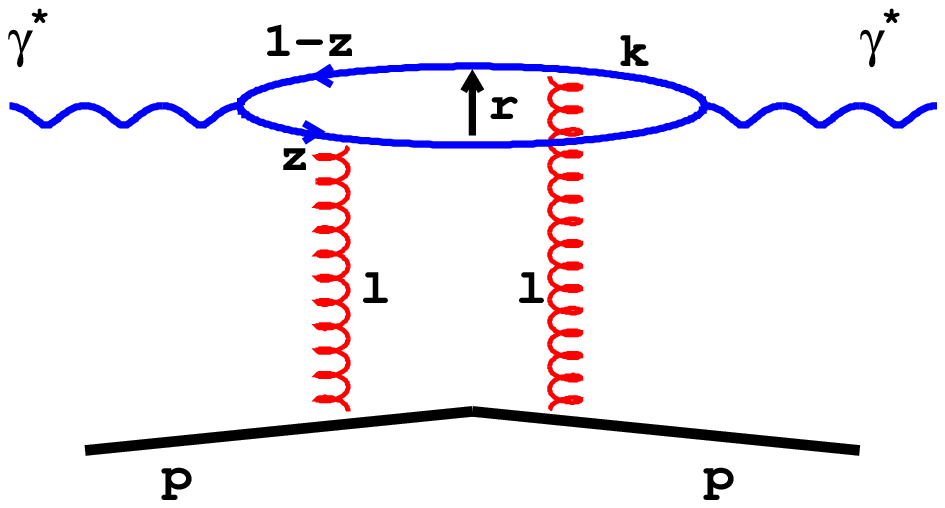} 
   \caption{Elastic photon-proton scattering at $t=0$.}
   \label{fig:elastic}
\end{figure}

We give here a derivation of the dipole representation for the elastic $\gamma^*p \rightarrow \gamma^* p$ scattering 
to exemplify the main properties of the dipole picture.
Dipoles incorporate naturally the interference effects between different quark-gluon couplings which are essential to obtain a proper description of DIS reactions at low $x$, namely colour transparency\footnote{Although we want to limit ourselves in this paper to $ep$ and $eA$ scattering only, let us note that the dipole concept, because of its importance in understanding the QCD evolution~\cite{Muell} and its success at HERA,  is now extensively investigated also in the context of pp scattering at the 
  LHC~\cite{LUND}.} .
For the sake of example let us consider the scattering of a longitudinally polarized  virtual photon on a proton.
Following the derivation given in~\cite{NZ} and~\cite{BGK,GB} we write the  total cross section in the $k_T$ factorization form according to the Feynman diagrams shown in Fig. 1 as  
 \begin{equation} 
\sigma_L = \frac{\alpha_{em}}{\pi}\sum_{f} e^2_f \int \frac{d^2 \vec{l}}{l^4}
\alpha_s f(x,l^2)\int d^2 \vec{k}\int dz
4Q^2z^2(1-z)^2  \left( \frac{1}{D(\vec{k})}-\frac{1}{D(\vec{k}+\vec{l})}  \right) 
\label{elastic}
\end{equation}
where $ D(\vec{k}) = \vec{k}^2 +\bar{Q}^2$ and  $\bar{Q}^2 = z(1-z)Q^2+ m^2_f$. $m_f$ denotes here the flavour dependent quark mass and $f(x,l^2)$ denotes the unintegrated gluon density.
In the $\gamma^*p$ collinear frame $\vec{k}$ and $\vec{l}$ are the two-dimensional transverse momentum vectors of the quark and the exchanged gluon and $z$ and $(1-z)$ are the fractions of the light-cone momentum of the photon carried by the quarks. 

In the elastic forward scattering there are just two possibilities for gluons to couple to the quarks shown by the two diagrams of 
Figure \ref{fig:elastic}.
Both diagrams  should be taken  into account because only then
the important property of  colour transparency  emerges from the cancellation between the two propagators in  eq.~\ref{elastic}.
In the limit $k\gg l$ the cancellation leads to $\sigma_L \approx 0$, reflecting the intuitively clear fact that a gluon cannot see a quark pair when its wave length is much larger than the distance between the quarks. 
In this case, the quark pair will appear as neutral to the gluon.    

Colour transparency emerges more naturally when the transverse quark momentum $\vec{k}$ is replaced by its  Fourier conjugate $\vec{r}$, the transverse separation between the two quarks. We have then
$$\int \frac{d^2\vec{k}}{2\pi} \exp(i\vec{k}\vec{r})\frac{1}{D(\vec{k})}=
K_0(r\cdot\bar{Q} )
$$ 
and 
$$
\left(\frac{1}{D(\vec{k})}-\frac{1}{D(\vec{k}+\vec{l})} \right)=\int\frac{d^2 \vec{r}}{2\pi} \exp(-i\vec{k}\vec{r}) (1-\exp(-i\vec{l}\vec{r})) 
K_0(r\cdot\bar{Q} )
$$
with $K_0$ being the Bessel-Mc Donald function.

The change of  variables from $\vec{k}$ to $\vec{r}$ leads to the dipole representation,
\begin{equation}  
\sigma_L ={\rm Im}A^{\gamma^* p}(x,Q,\vec{\Delta}=0)=\int d^2\vec{r}\int dz \sum_f \Psi_L^*(r,z,Q^2) \sigma_{qq}(x,r) \Psi_L(r,z,Q^2)
\label{xtot}
\end{equation}
in which the wave function $\Psi_L$ describe the probability amplitude to find a $q\bar q$ pair within a virtual incoming or outgoing photon.   
$$
\Psi_L^2=\frac{3\alpha_{em}}{2\pi^2}e_f^2 4 Q^2 z^2(1-z)^2K_0^2(r\cdot \bar{Q})
$$
The dipole cross section $\sigma_{qq}$ describe the interaction of the $q\bar q$ pair with the proton mediated by the two gluon exchange.
$$
\sigma_{qq}=\frac{2\pi}{3}\int\frac{d^2\vec{l}}{l^4} \alpha_s f(x,l^2)
(1-e^{-i\vec{l}\vec{r}}) (1-e^{i\vec{l}\vec{r}}) 
=\frac{4\pi^2}{3}\int\frac{d l^2}{l^4} \alpha_s f(x,l^2)(1-J_0(lr))
$$ 
By introducing the relation between the integrated and unintegrated gluon density \begin{equation}
xg(x,Q^2)=\int_0^{Q^2} dl^2 f(x,l^2)/l^2,
\end{equation}
and assuming that $\alpha_s$ depends on the dipole size only, 
the dipole cross section can be further simplified by approximating $(1-J_0(lr))\approx (lr)^2/4$, which is valid for $l^2<1/r^2$
\begin{equation}  
\sigma_{qq}=\frac{\pi^2}{3}\alpha_s(1/r^2)r^2xg(x,1/r^2).
\label{xdipole}
\end{equation}  
Here $xg(x,\mu^2)$ is the gluon distribution which evolves in $\mu^2=1/r^2$ according to the DGLAP evolution equation.

In this representation  colour transparency becomes a property of the dipole cross section, for small dipoles $r\rightarrow 0$ also $\sigma_{qq} \rightarrow 0$, as is intuitively clear. The last form of the dipole cross section was first derived  in ref~\cite{FRS}, in an alternative way.

\subsection{Woods-Saxon Distribution}
\begin{figure}[t]
  \centering
  \includegraphics[width=0.55\textwidth]{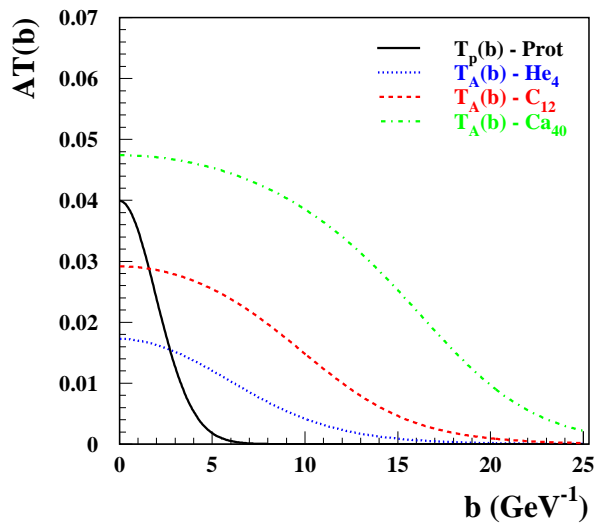}
\caption{\label{fig:tfunc}
  The transverse density $A\,T_{WS}(b)$  for
  several light nuclei compared to the  proton transverse profile, $T_p(b)$. 
}
\end{figure}

The distribution of nucleons in the nucleus $\rho_{A}(r)$  
is usually parametrized according to the Woods-Saxon distribution \cite{BohrMot}
\begin{equation}
   \rho_{WS}(r) = \frac{N}{
                           \exp\left( \frac{(r-R_A)
                                           }{ \delta
                                           }
                               \right)
                      + 1 } \;,
\end{equation}
with  $\delta =
0.54 \, \mbox{fm}$, $R_A = (1.12\, \mbox{fm})\, A^{1/3}  - 
(0.86\, \mbox{fm})\, A^{-1/3}$ 
and $N$ is adjusted to normalize the distribution to one
\begin{equation}
   \int d^3\vec{r} \,\rho_{WS}(r) = 1 \; .
\end{equation}
The transverse distribution is defined as
\begin{eqnarray}
T_{WS}({ b}) = \int_{-\infty}^{+\infty} dz \rho_{WS} (\sqrt{ { b^2} +
z^2} ).
\end{eqnarray}
Figure~\ref{fig:tfunc} compares the proton shape $T_p(b)$ with
the transverse density $T_{WS}(b)$ 
 for several light nuclei.

\end{document}